\documentstyle[prd,aps,epsfig]{revtex}
\tighten

\newcommand{\IE}{{i.e.}}
\newcommand{\EG}{{e.g.}}
\newcommand{\ETAL}{{\it et al.}}
\newcommand{\PREQ}[1]{[{\tt #1}]}
\newcommand{\PREP}[1]{, [{\tt #1}]}
\newcommand{\BIBT}[1]{``#1'', }

\newcommand{\UUNIT}[2]{\,{\mbox{#1}^{#2}}}
\newcommand{\SK}[1]{\mbox{s}_{#1}}

\newcommand{\SS}[1]{{\scriptscriptstyle{#1}}}

\newcommand{\PL}{\SS{\rm Pl}}

\newcommand{\START}{\SS{\rm start}}
\newcommand{\CRIT}{\SS{\rm crit}}
\newcommand{\QUINT}{\SS{\rm Q}}
\newcommand{\XI}{\SS{\rm \xi}}
\newcommand{\KAPPA}{\SS{\rm \kappa}}
\newcommand{\LEFT}{\SS{\rm L}}
\newcommand{\RIGHT}{\SS{\rm R}}

\newcommand{\PN}{\SS{\rm PN}}
\newcommand{\DN}{\SS{\rm DN}}
\newcommand{\EX}{\SS{\rm EX}}

\newcommand{\GRAV}{\SS{\rm g}}
\newcommand{\COUPL}{\SS{\rm co}}

\newcommand{\MAT}{\SS{\rm mat}}
\newcommand{\RAD}{\SS{\rm rad}}
\newcommand{\COURB}{\SS{K}}

\newcommand{\NUC}{\SS{\rm nuc}}

\newcommand{\ANG}{\SS{\rm A}}
\newcommand{\LUM}{\SS{\rm L}}

\newcommand{\ENV}{\SS{\rm env}}
\newcommand{\DIFF}{\SS{\rm diff}}
\newcommand{\THERM}{\SS{\rm th}}
\newcommand{\THM}{\SS{\rm Th}}
\newcommand{\CHAND}{\SS{\rm Ch}}
\newcommand{\MAX}{\SS{\rm max}}
\newcommand{\EXP}{\SS{\rm exp}}
\newcommand{\GR}{\SS{\rm GR}}
\newcommand{\PEAK}{\SS{\rm P}}

\newcommand{\EFF}{\SS{\rm eff}}

\newcommand{\MC}{\SS{\rm MC}}

\newcommand{\NEWT}{{\rm N}}
\newcommand{\Hconf}{{\cal H}}
\newcommand{\LAGR}{{\cal L}}
\newcommand{\OO}{{\cal O}}
\newcommand{\dd}{{\rm d}}

\newcommand{\FER}{{\rm Fe}}
\newcommand{\COB}{{\rm Co}}
\newcommand{\NIC}{{\rm Ni}}

\newcommand{\ATOME}[5]
{{}^{#1}_{#2}\mathrm{#3}^{#4}_{#5}}

\newcommand{\Chi}{{\rm X}}
\newcommand{\VV}[1]{\overline #1}
\newcommand{\TT}[1]{\overline {\VV #1}}

\newcommand{\EQN}[1]{Eq.~{#1}}
\newcommand{\EQNS}[1]{Eqs.~{#1}}
\newcommand{\EEQN}[1]{Equation~{#1}}

\begin{document}
\draft

\title{Cosmological observations in scalar-tensor quintessence}

\author{Alain Riazuelo} 

\address{Service de Physique Th\'eorique, CEA/DSM/SPhT Unit\'e de
recherche associ\'ee au CNRS CEA/Saclay, \\ F--91191 Gif-sur-Yvette
c\'edex (France), \\ Universit\'e de Gen\`eve, D\'epartement de
Physique Th\'eorique, 24, quai Ernest Ansermet, CH--1211 Gen\`eve,
Switzerland,}

\vskip0.25cm 

\author{Jean--Philippe Uzan} 

\address{Laboratoire de Physique Th\'eorique, CNRS-UMR 8627, \\ B\^at.
  210, Universit\'e Paris XI, F--91405 Orsay c\'edex (France), \\
  Institut d'Astrophysique de Paris, 98bis boulevard Arago, F--75014
  Paris, France}

\date{9 June 2002}

\maketitle

\begin{abstract}
  The framework for considering the astronomical and cosmological
  observations in the context of scalar-tensor quintessence in which
  the quintessence field also accounts for a time dependence of the
  gravitational constant is developed.  The constraints arising from
  nucleosynthesis, the variation of the constant, and the
  post-Newtonian measurements are taken into account. A simple model
  of supernovae is presented in order to extract the dependence of
  their light curves with the gravitational constant; this implies a
  correction when fitting the luminosity distance.  The properties of
  perturbations as well as CMB anisotropies are also investigated.
\end{abstract}


\pacs{{\bf preprint:} SPhT-Saclay T02/011}

\section{Introduction}

The combination of recent astrophysical and cosmological observations
(among which are the luminosity distance versus redshift relation up
to $z \sim 1$ from type Ia supernovae~\cite{sndata}, the cosmic
microwave background temperature anisotropies~\cite{cmbdata} and
gravitational lensing~\cite{gldata}) seems to indicate that the
universe is accelerating and thus that about $70\%$ of the energy
density of the universe is made of a matter with a negative pressure
(\IE, having an equation of state $\omega \equiv P / \rho < 0$).  This
raises the natural question of the physical nature of this matter
component. Indeed, a solution would be to have a cosmological constant
(for which $\omega = - 1$) but one will then have to face the well
known {\it cosmological constant problem}~\cite{weinberg}, \IE, the
fact that the value of this cosmological constant inferred from the
cosmological observation is extremely small --- about 120 order of
magnitude --- compared with the energy scales of high energy physics
(Planck, grand unified theory, strong and even electroweak
scales). Another solution is to argue that there exists a (yet
unknown) mechanism which makes the cosmological constant strictly
vanish and to find another matter candidate able to explain the
cosmological observations. Indeed it assumes that the cosmological
constant problem is somehow solved and replaces it by a {\it dark
energy problem}.

This latter route has gained a lot of enthusiasm in the past years and
many candidates have been proposed (for recent reviews on see, \EG,
Refs.~\cite{binetruy00,carroll00}). Among all these proposals,
quintessence~\cite{ratra,wett,cds} seems to be the most promising
mechanism. In these models, a scalar field is rolling down a potential
decreasing to zero at infinity (often referred to as a runaway
potential) hence acting as a fluid with an equation of state varying
in the range $- 1 \leq \omega < 1$ if the field is minimally coupled.
Runaway potentials such as the exponential potential and inverse power
law potentials
\begin{equation}
\label{01}
V (\phi) = \frac{M^{4 + \alpha}}{\phi^\alpha} ,
\end{equation}
with $\alpha > 0$ and $M$ a mass scale, can be found in models where
supersymmetry is dynamically broken~\cite{binetruy99} and in which
flat directions are lifted by nonperturbative effects.

As clearly explained in Ref.~\cite{brax99}, all the models for this
dark energy have to (i) show that they do not contain in a disguise
way a cosmological constantlike fine-tuning (the {\em fine-tuning
problem}), (ii) explain why this kind of matter starts to dominate
today (the {\it coincidence problem}), (iii) give an equation of state
compatible with the observational data (the {\it equation of state
problem}) and (iv) arise from some high energy physics mechanisms (the
{\it model building problem}).  Quintessence models mainly solved the
fine-tuning problem because of the existence of tracking
solutions~\cite{zlatev} (first studied in Ref.~\cite{ratra,wett})
which are scaling attractor solutions of the field equations and
allows the initial conditions for the scalar field to vary by about
150 orders of magnitude. The second tuning (related to the
coincidence) concerns the mass scale $M$ that has to be determined by
the requirement that about $70\%$ of the energy density of the
universe is in the quintessence field. As shown in
Ref.~\cite{riazuelo00} for the case of the inverse power law
potential, this mass scale is comparable to other scales from high
energy physics and the tuning on this mass scale is mild provided the
exponent $\alpha$ is not too small ($\alpha$ must be bigger than $4$
so that $M > 1 \UUNIT{TeV}{}$). The equation of state depends on the
shape of the potential and it is hoped that it will be soon determined
by, \EG, a weak lensing experiment~\cite{bb}. For instance, it has
been shown that an exponential potential cannot lead to an
accelerating universe~\cite{ratra,wett} and that the equation of state
for an inverse power law potential mainly depends on the slope
$\alpha$.  As explained above, the model building is also addressed in
the framework of supersymmetry

Quintessence scenarios, however, have some important problems. The
requirement of slow roll (mandatory to have a negative pressure) and
the fact that the quintessence field dominates today imply that (i) it
is very light~\cite{carollprl} (roughly of the order $\sim 10^{- 33}
\UUNIT{eV}{}$ and it should induce violation of the equivalence
principle and time variation of the gravitational constant) and that
(ii) the vacuum expectation value of the quintessence field today is
of order of the Planck mass.  This latter problem led Brax and
Martin~\cite{brax99,brax1} to propose that a supergravity correction
had to be taken into account leading to the so-called supergravity
(SUGRA) quintessence potential
\begin{equation}
\label{sugra}
V(\phi)
 = \frac{M^{4 + \alpha}}{\phi^\alpha} 
   e^{\frac{1}{2}{\phi^2}/{M_\PL^2}} ,
\end{equation}
which shares the same properties as the inverse power law
potential at early time but which stabilizes the quintessence
field accounting for a better agreement of the equation of state.

Note that the two main features of a quintessence potential is that it
must be steep enough for the field to be in a kinetic regime for a
large set of initial conditions, hence redshifting faster than
radiation and being subdominant at nucleosynthesis, and then to reach
a slow roll regime to mimic a cosmological constant.  This latter
regime will always ultimately take place and the parameters of the
potential have to be tuned so that it happens around today.  The
simplest waya to implement this idea are to use inverse power law
potentials and exponential potentials which are one parameter
potentials. Another solution (but involving more parameters) is to
consider potentials with a local minimum as was first proposed by
Wetterich~\cite{wett}.

An underlying motivation to replace the cosmological constant by a
time dependent scalar field probably lies in string models in which
any dimensionful parameter is expressed in terms of the fundamental
string mass scale and the vacuum expectation value of a scalar field.
For instance, string theory has revived the consideration of
gravitational-strength scalar fields~\cite{gsw} such as Kaluza-Klein
moduli or the dilaton appearing in the low energy limit of the
gravitational sector leading to scalar-tensor theories of gravity. As
explained above, the quintessence field is expected to be very light
and this points toward scalar-tensor theories of gravity in which a
light (or massless) scalar field can be present in the gravitational
sector without being phenomenologically disastrous. These arguments
lead to consider quintessence models in the framework of scalar-tensor
gravity. Indeed, the dilaton and the quintessence field can be two
different scalar fields (as considered, \EG, in Ref.~\cite{bp} in the
particular case of Brans-Dicke theory) or the same scalar field.  The
latter subclass involving a single scalar field (the quintessence
field is also the dilaton) dictating the time variation of both the
gravitational constant and the cosmological constant is attractive; it
involves less free functions and has been focused on much in recent
years. The study of these quintessence models, referred to as
nonminimal quintessence~\cite{uzan99}, coupled
quintessence~\cite{amendola99b}, extended
quintessence~\cite{perrotta99} or generalized
quintessence~\cite{gef00b} was mainly motivated by the fact that
tracking solutions have been shown to exist for nonminimally coupled
scalar fields~\cite{uzan99,amendola99a}.

Scalar-tensor theories are the most natural extensions of general
relativity, in particular they contain local Lorentz invariance,
constancy of nongravitational constants and respect the weak
equivalence principle. The most general action for these
theories~\cite{will} for the matter and gravity is given, in the
Jordan frame, by
\begin{equation}
\label{1}
S = \int \left[- F (R, \phi)
               + \frac{1}{2} \frac{\omega(\phi)}{\phi}
                 \partial_\mu \phi \partial^\mu \phi
               - V (\phi)
               + \LAGR_\MAT \right] \sqrt{-g} \dd^4 x ,
\end{equation}
where $\LAGR_\MAT$ is the Lagrangian of ordinary matter (such as
radiation and pressureless matter), $g$ is the determinant of the
metric $g_{\mu\nu}$, $R$ the Ricci scalar, and $V$ is a potential to
be discussed below. The action~(\ref{1}) depends {\it a priori} on
three arbitrary functions $F$, $\omega$, and $V$. If $F (R, \phi)$ is
not a trivial function of $R$ then one has an additional (massive)
scalar degree of freedom~\cite{wands94} so that the requirement that
there is only one scalar partner to the graviton implies
\begin{equation}
\label{2}
F (R, \phi) = \frac{1}{2 \kappa} F(\phi) R ,
\end{equation}
where $\kappa \equiv 8 \pi G \equiv M_\PL^2$, $G$ being the bare
Newton constant.  $F(\phi)$ is a dimensionless function which needs to
be positive to ensure that the graviton carries positive
energy~\cite{damour92}.  Now, by a redefinition of $\phi$ we can
always choose either $F (\phi)$ or $\omega (\phi) / \phi$ to be unity
so that we are left with only two independent functions of the scalar
field one being the potential $V$. Choosing $\omega (\phi) / \phi =
1$, the effective (time dependent) constant is related to $G$ by
\begin{equation}
\label{4}
G_\EFF (\phi) \equiv \frac{G} {F (\phi)} .
\end{equation}

As was explained above, authors first considered models with a
nonminimally coupled scalar
field~\cite{uzan99,amendola99a,ritis99,bertolami99},
\IE, in which the function $F$ is decomposed as
\begin{equation}
\label{3}
F (\phi) = 1 + 2 \xi \kappa f(\phi) .
\end{equation}
Indeed, as long as we have not fixed the choice of the function $f$,
such a decomposition is completely general and always possible. If the
normalization of $f$ is chosen in a way that $f (M_\PL) \sim \OO
(M_\PL^2)$ today then $\xi$ gives an order of magnitude of the
deviation with respect to general relativity today.

Due to the time variation of the gravitational constant, the strength
of the coupling $\xi$ can be constrained once the function $f (\phi)$
has be chosen. Chiba~\cite{chiba99} gave the constraints arising from
the post-Newtonian (PN) parameters and the time variation of $G$ for
$f = \phi^2 / 2$. In that case the deviation of the scalar-tensor
theory from general relativity was fixed but, as pointed out by
Bartolo and Pietroni~\cite{bartolo} in scalar-tensor quintessence
models a ``double attractor mechanism'' can happen, namely, of the
scalar-tensor theory towards general relativity (through the
Damour-Nordtvedt mechanism~\cite{DM}) and of the quintessence field
toward its tracking solution hence allowing for large deviation from
general relativity in the early universe. Note that in these models
the dilaton is not completely stabilized and is slow rolling in its
runaway potential.

Many works have then studied the cosmological implications of these
models, starting from the study of the perturbations in Brans-Dicke
theory~\cite{chenk} and for a nonminimally coupled scalar
field~\cite{bacci,la} the computation of cosmic microwave background
(CMB) anisotropies~\cite{perrotta99,bommamendola}, the properties of
the nucleosynthesis~\cite{bartolo,chen00}. But among all, one of the
very interesting results concerns the possibility to rule out some of
these models~\cite{gef00b}: it was shown that because of the
positivity of the energy of the graviton, one can mimic a model in
which the gravity is described by general relativity with a
cosmological constant by a scalar-tensor theory with $V = 0$ (or when
certain relation between $V$ and $F$ are set) only in a certain range
of redshift, hence offering a powerful test on a large class of
scalar-tensor models.

In this article, we try to extract the observational constraints on
these quintessence models in scalar-tensor gravity. We first recall in
Sec.~\ref{II} the background properties: we study the constraints
arising from the bounds on the post-Newtonian parameters, the time
variation of the gravitational constant, of nucleosynthesis and of the
positivity of the energy of the graviton. We then turn to the
information that can be extracted for the type Ia supernovae
(Sec.~\ref{IIB}). This lead us to describe a simple model of SN~Ia in
order to extract the dependence of the light curve with the
gravitational constant and give us a modified magnitude vs redshift
relation that has to be used when extracting the luminosity distance
vs redshift relation. We finish by a computation of the angular
diameter distance vs redshift relation (Sec.~\ref{IIC}). We then turn
to the property of the perturbations and generalize the attractor
property (Sec.~\ref{IV}) found in Ref.~\cite{brax00} to the
scalar-tensor quintessence and then discuss the CMB angular spectrum
(Sec.~\ref{V}).  All the technical details are gathered in the
appendices and we apply these results all along the article to two
examples, one being a nonminimally coupled scalar field and the second
a scalar-tensor theory with conformal coupling.

\section{Background properties}
\label{II}

We consider a Friedmann-Lema\^{\i}tre universe with the line element
\begin{equation}
\label{5}
\dd s^2 = \dd t^2 - a^2 (t) \gamma_{ij} \dd x^i \dd x^j
        \equiv g_{\mu\nu} \dd x^\mu \dd x^\nu ,
\end{equation}
where $a$ is the scale factor, $t$ the cosmic time and $\gamma_{ij}$
the metric on constant time hypersurfaces. Greek indices run from 0 to
3 and Latin indices from 1 to 3. As detailed in Appendix~\ref{B}, the
Friedmann equation in presence of a nonminimally coupled scalar field
takes the form
\begin{equation}
\label{6}
\Hconf^2 + K
 = \frac{\kappa_\EFF}{3} a^2
   \left(\rho_\MAT + \rho_\MC + \rho_\XI \right) ,
\end{equation}
where we have introduced the conformal Hubble parameter $\Hconf \equiv
\dot a/a$, and where an overdot denotes a derivative with respect to the
conformal time ($\dd t = a \dd \eta$), and $K$ is the curvature
index. On the right-hand side, $\rho_\MAT$ designs the matter energy
density, and $\rho_\MC$ and $\rho_\XI$, the energy densities of the
scalar field and respectively defined respectively in Eqs.~(\ref{B3})
and~(\ref{B5}). The quantity $\kappa_\EFF \equiv \kappa / F (\phi)$
acts as an effective Newton constant and depends also on $\phi$ (and
hence, varies with time).  We define the density parameter of any
component $f$ by
\begin{equation}
\label{4a}
\Omega_\SS{f} \equiv \frac{\kappa_\EFF a^2 \rho_\SS{f}} {3 \Hconf^2} ,
\end{equation}
where one has to use $\kappa_\EFF$ and not $\kappa$ since we have only
access to a measure of the effective Newton constant and not of its
bare value.

\subsection{How is $F (\phi)$ constrained?}
\label{IIa}

In this section, we review the different constraints on $F (\phi)$
arising from the requirement of the positivity of the energy of the
graviton, the bounds on the post-Newtonian parameters and of the time
variation of the gravitational constant and from nucleosynthesis.

\begin{enumerate}

\item\underline{Positivity of the energy of the graviton}:
  Diagonalizing the action~(\ref{1}) by the conformal
  transformation~\cite{damour92}
\begin{equation}\label{conf}
\widetilde g_{\mu\nu} \equiv F(\phi) g_{\mu\nu}
,\quad
\left( \frac{\dd \widetilde\phi}{\dd\phi} \right)^2
  \equiv   \frac{3}{4} \left( \frac{\dd \ln F (\phi)}{\dd \phi} \right)^2
         + \frac{1}{2 F (\phi)}
,\quad 
A (\widetilde \phi) \equiv F^{- 1 / 2}(\phi)
,\quad
2 U (\widetilde \phi) \equiv \frac{V (\phi)}{F^2 (\phi)} ,
\end{equation}
  one can show that the graviton is the perturbation of $\widetilde
  g_{\mu\nu}$ and its scalar partner $\widetilde \phi$ evolving in the
  potential $U(\widetilde \phi)$. It can be shown that a scalar-tensor
  theory in fact well defined only if the transformation~(\ref{conf})
  is possible~\cite{gef00b,damour92}.  This is the case if $F (\phi)$
  is positive. When integrating our equations this constraint has to
  be checked. In practice, since one has $\kappa_\EFF \propto F^{-1}
  \propto \Hconf^2 + K$, this constraint is always satisfied when the
  universe is spatially flat.

\item\underline{Post-Newtonian constraints}: 
  With the form~(\ref{1}), the standard post-Newtonian parameters are
  given by~\cite{will,damour92} 
\begin{eqnarray} 
\label{8} 
\gamma_\PN - 1
 & = & \frac{\left(F'_0\right)^2} { M_\PL^2 F_0 + 2 \left(F'_0\right)^2}
   =   - 2 \frac{\alpha_0^2}{1 + \alpha_0^2} , \\
\label{9} 
\beta_\PN - 1
 & = & \frac{1}{4} \frac{F_0 F'_0} {2 M_\PL^2 F_0 + 3 \left(F'_0\right)^2} 
                   \frac{\dd \gamma_\PN}{\dd \phi}
   =   \frac{1}{2} \frac{\alpha_0^2}{(1 + \alpha_0^2)^2} 
                   \frac{\dd \alpha_0}{\dd \widetilde \phi} , 
\end{eqnarray} 
  where a dash denotes a derivative with respect to the field $\phi$
  and where a subscript $0$ means that the function is evaluated
  today.  The function $\alpha$ is defined by 
\begin{equation}
\label{coupling}
\alpha (\widetilde \phi) \equiv \frac{\dd \ln A}{\dd \widetilde \phi} .  
\end{equation} 
  Current constraints (see, \EG, Ref.~\cite{td} for a recent review of
  the measurements) give
\begin{equation} 
\left|\gamma_\PN - 1\right| \leq 2 \times 10^{- 3} 
,\quad 
\left|\beta_\PN - 1\right| \leq 6 \times 10^{- 4} .  
\end{equation} 
  This implies that 
\begin{equation}
\alpha_0^2 \sim \frac{1}{M_\PL^2 F_0} \left(F'_0\right)^2 < 10^{- 3} ,
\end{equation} 
  and, as explained in Ref.~\cite{gef00b}, the second bound cannot be
  used to constraint $\dd \alpha_0 / \dd \phi$.

\item\underline{Time variation of $G$}: The effective constant
  $G_\EFF$ deduced from Eq.~(\ref{4}) is not the gravitational
  constant that would be measured in a Cavendish-Michel-type
  experiment, \IE, it is not the effective Newton constant. This
  constant $G_{\NEWT_\EFF}$, entering in the force between two masses,
  is given by
\begin{equation}
G_{\NEWT_\EFF} = \frac{G}{F_0} 
                 \frac{2 M_\PL^2 F_0 + 4 \left(F'_0\right)^2} 
                      {2 M_\PL^2 F_0 + 3 \left(F'_0\right)_0^2}
               = G A_0^2 (1 + \alpha_0^2) , 
\end{equation}
  in which one has two contributions, namely the exchange of a
  graviton and of a scalar~\cite{damour92}. Current
  constraints~\cite{dickey} on the variation of the Newton constant
  imply 
\begin{equation}
\left|\frac{\dot G_{\NEWT_\EFF}}{G_{\NEWT_\EFF}}\right|
 \leq 6 \times 10^{- 12} \UUNIT{yr}{-1} .
\end{equation}
  
\item\underline{Nucleosynthesis}: Nucleosynthesis bounds have two
  origins.  Roughly speaking, we have to require that (i) the matter
  contents dominating the Friedmann equation behaves as radiation at
  nucleosynthesis and that (ii) the effective number of degrees of
  freedom of the relativistic particles, $g_*$ say, does not vary from
  more than $20\%$ than its expected value $g_* = 10.75$ at this
  epoch.
  
  In the case of quintessence with an exponential potential, these
  bounds were used to show that the quintessence field can not close
  the universe~\cite{fj}. For general inverse power law and SUGRA
  potentials, the first constraint was shown~\cite{riazuelo00} to
  imply bounds on the initial condition of the scalar field at the end
  of inflation. Here, we want to estimate the bounds arising from the
  second requirement and we assume that the contribution of the scalar
  field is subdominant with respect to the radiation. The energy
  density of the radiation is given by
\begin{equation}
\rho_\RAD = \frac{\pi^2}{30} g_* T^4 , 
\end{equation} 
  where $T$ is the temperature. Assuming that $\rho_\QUINT \ll
  \rho_\RAD$ [$\rho_\QUINT$ being defined by Eq.~(\ref{B7})], the
  Friedmann equation~(\ref{6}) leads to
\begin{equation} 
\Hconf^2 = \kappa_0 a^2 \frac{\pi^2}{90} 
           g_* \left(1 + \frac{\delta g_*}{g_*} \right) T^4 , 
\end{equation} 
  with
\begin{equation}
\frac{\delta g_*}{g_*}
 \equiv \left( \frac{\kappa_\EFF}{\kappa_0} - 1 \right)
 =      \frac{F (\phi_0) - F(\phi_\NUC)}{F(\phi_\NUC)} ,
\end{equation}
  where $\phi_\NUC$ is the value of $\phi$ at the time of
  nucleosynthesis.  Nucleosynthesis therefore imposes that
\begin{equation} 
\label{bbn} 
\left|\frac{\delta g_*}{g_*}\right| \leq 0.2
 \Longleftrightarrow
0.8 \leq   \left| \frac{F (\phi_0)}{F (\phi_\NUC)}\right| 
         = \left| \frac{A^2 (\phi_\NUC)}{A^2 (\phi_0)}\right| \leq 1.2 .  
\end{equation} 
  This constraint, in the framework of quintessence, was first derived
  by Bartolo and Pietroni~\cite{bartolo}. It was also
  shown~\cite{chen00} numerically that in some nonminimally coupled
  quintessence models the Helium abundance can be reduced extending
  the upper limit on the number of neutrino to 5. Let us also
  emphasize that very large values of $|{F (\phi_0)} / {F
  (\phi_\NUC)}|$ were shown~\cite{pichondamour} to be consistent with
  the observed abundances of light elements if $\dd^2 A / \dd
  \widetilde \phi^2$ is large enough. Hence, the naive
  limit~(\ref{bbn}) can be much more stringent than a detailed
  (numerical) study may show.
\end{enumerate}

\subsubsection{Nonminimally coupled scalar field}

As a first example, we consider the case of the nonminimally scalar
field introduced in Ref.~\cite{uzan99} for which
\begin{equation}
F(\phi) = 1 + 2 \xi \kappa f(\phi) ,
\end{equation}
and we choose the function $f$ to be
\begin{equation}
\label{coupling2}
f(\phi) = \frac{1}{2} \phi^2 .
\end{equation}

\begin{figure}[ht]
\centerline{
\psfig{file=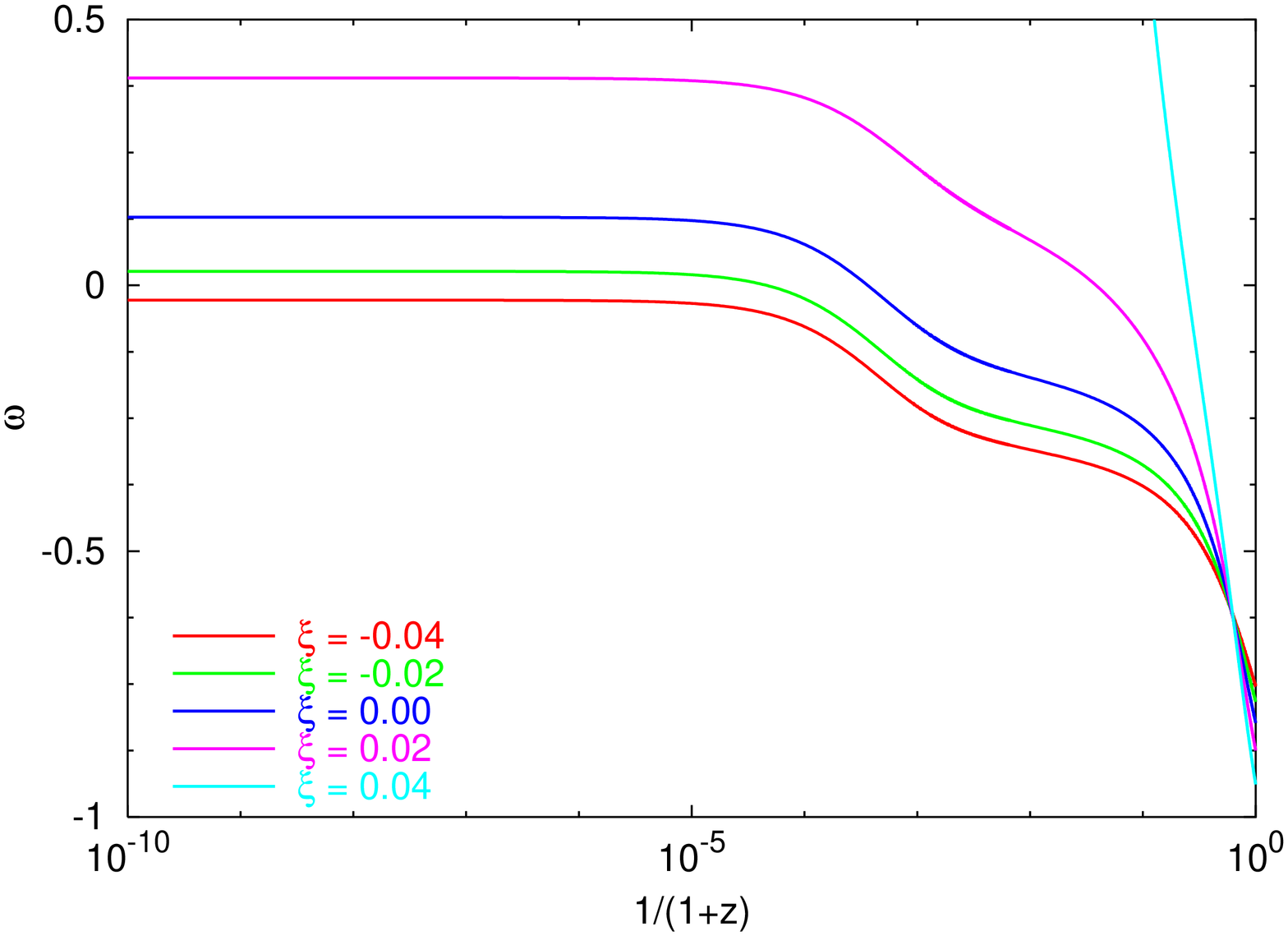,angle=270,width=3.5in}
\psfig{file=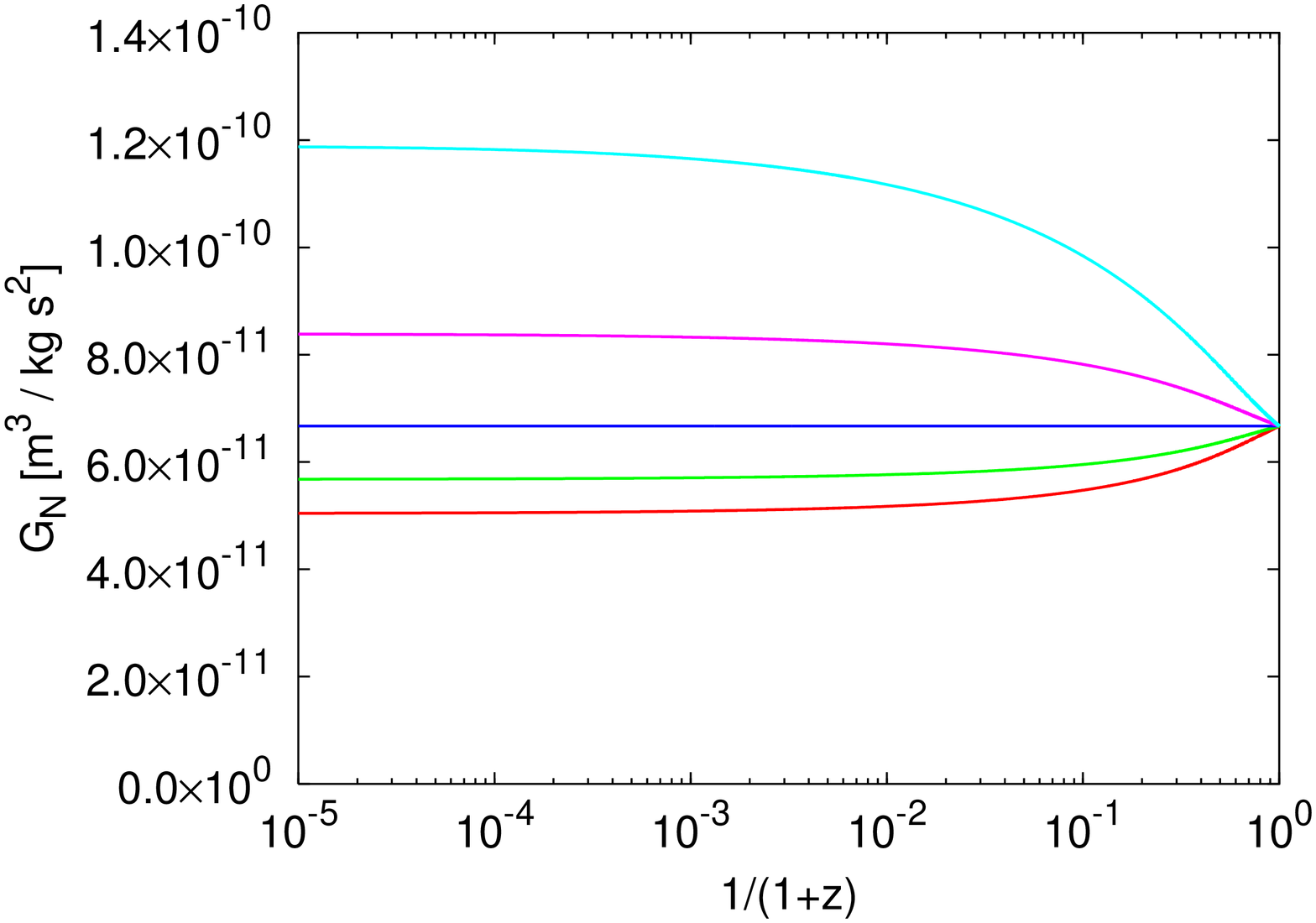,angle=270,width=3.5in}}
\caption{Evolution of the quintessence equation of state
  parameter $\omega_\QUINT$ as a function of the redshift $z$ for
  $\Omega_\QUINT = 0.7$ and various values of the coupling parameter
  $\xi$ (left). For negative $\xi$, $-\frac{1}{3} < \omega_\QUINT <
  \omega_{\rm B}$ in the tracking regime, whereas this condition does
  not hold for positive $\xi$.  This is due to the fact that
  $\rho_\XI$ is negative when $\xi$ is positive. For all the models
  presented here, $\omega_\QUINT -\left.\omega_\QUINT\right|_{\xi =
  0}$ changes sign around $z \simeq 1$.  Evolution of the
  gravitational constant as a function of the redshift $z$ for various
  values of the coupling parameter $\xi$ (right). The color codes are
  identical in the two plots.}
\label{fig0}
\end{figure}

In Fig.~\ref{fig0}, we depict the typical evolution of the equation of
state parameter $\omega_\QUINT \equiv P_\QUINT / \rho_\QUINT$ of the
quintessence field and the time variation of the Newton constant in
the minimally coupled and nonminimally coupled cases. As explained in
the appendices, the equation of state parameter of the quintessence
field has two distinct contributions: $\omega_\MC$ that appears in the
minimally coupled case, and an extra contribution $\omega_\XI$ that
appears only in the non minimally coupled case. $\omega_\QUINT$,
$\omega_\MC$, and $\omega_\XI$ are related by
\begin{equation}
\omega_\QUINT
 =   \frac{\rho_\MC}{\rho_\MC + \rho_\XI} \omega_\MC
   + \frac{\rho_\XI}{\rho_\MC + \rho_\XI} \omega_\XI .
\end{equation}
It can be trivially checked that when the tracking solution is reached,
\begin{equation}
\omega_\XI = - \frac{1}{3} .
\end{equation}
Now, a subtlety arises from the fact that depending on the sign of
$\xi$, the quantity $\rho_\XI$ can be negative. In this case,
$\omega_\QUINT$ does not lie between $\omega_\XI$ and
$\omega_\MC$. With our conventions, $\rho_\XI$ and $\xi$ are of
opposite sign.  In both cases one finds, however, that the field
reaches Planck values as it starts to dominate, \IE, today. Given the
form of the coupling function~(\ref{coupling2}), this means that all
the departure from general relativity are mostly felt today. This also
implies that these departures are stronger in the case of an inverse
power-law than in the case of a SUGRA potential, mainly because in the
latter case the potential is less steep (due to the exponential term)
so that the scalar field is more stabilized and rolls down slower,
implying a smaller time variation of the coupling function $F$.
Another interesting feature lies in the fact that $\omega_\QUINT
-\left.\omega_\QUINT\right|_{\xi = 0}$ changes sign around $z \simeq
1$.

\begin{figure}[ht]
\centerline{
\psfig{file=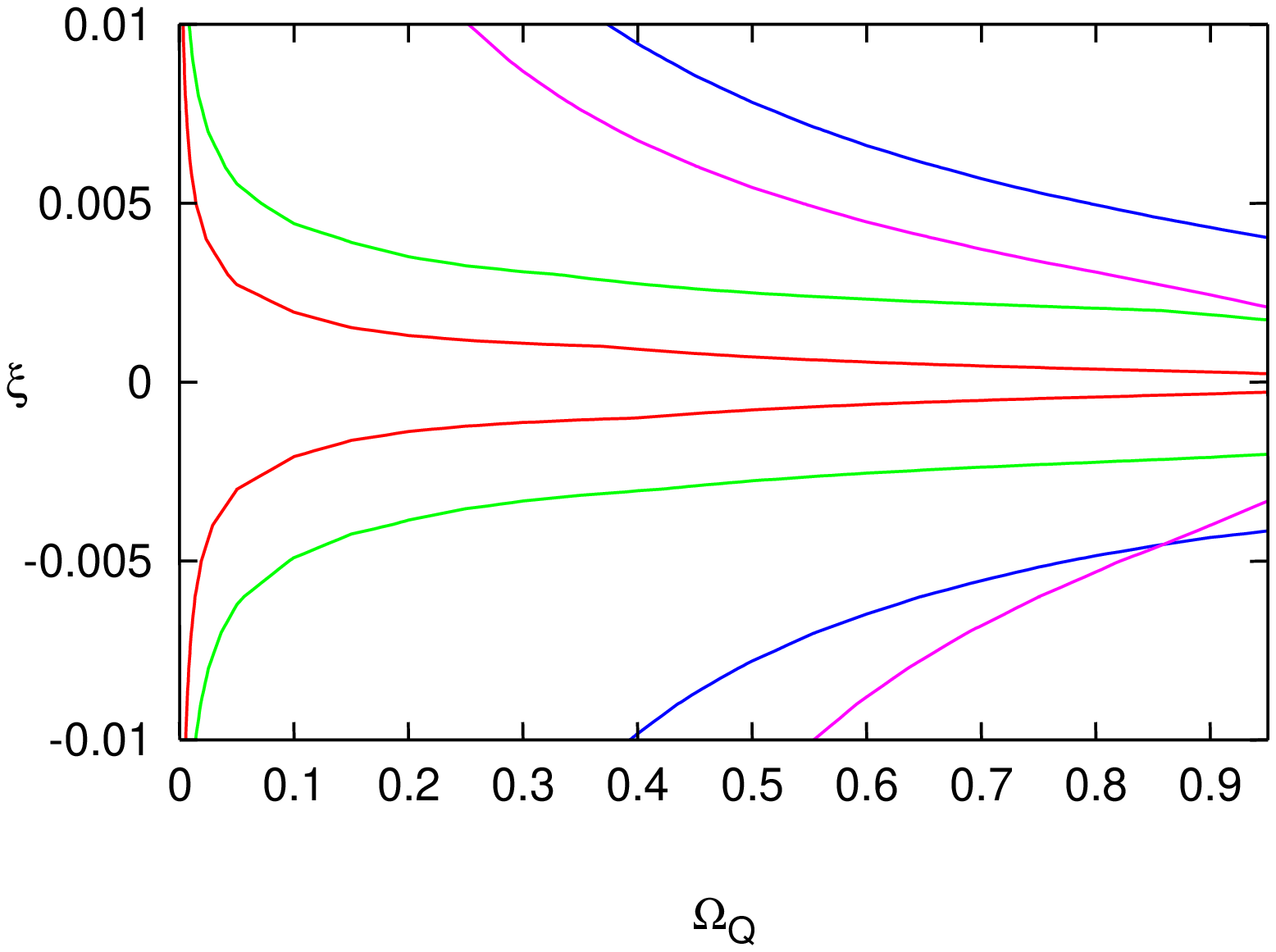,bbllx=135pt,bblly=150pt,
       bburx=500pt,bbury=630pt,angle=270,width=3.5in}
\psfig{file=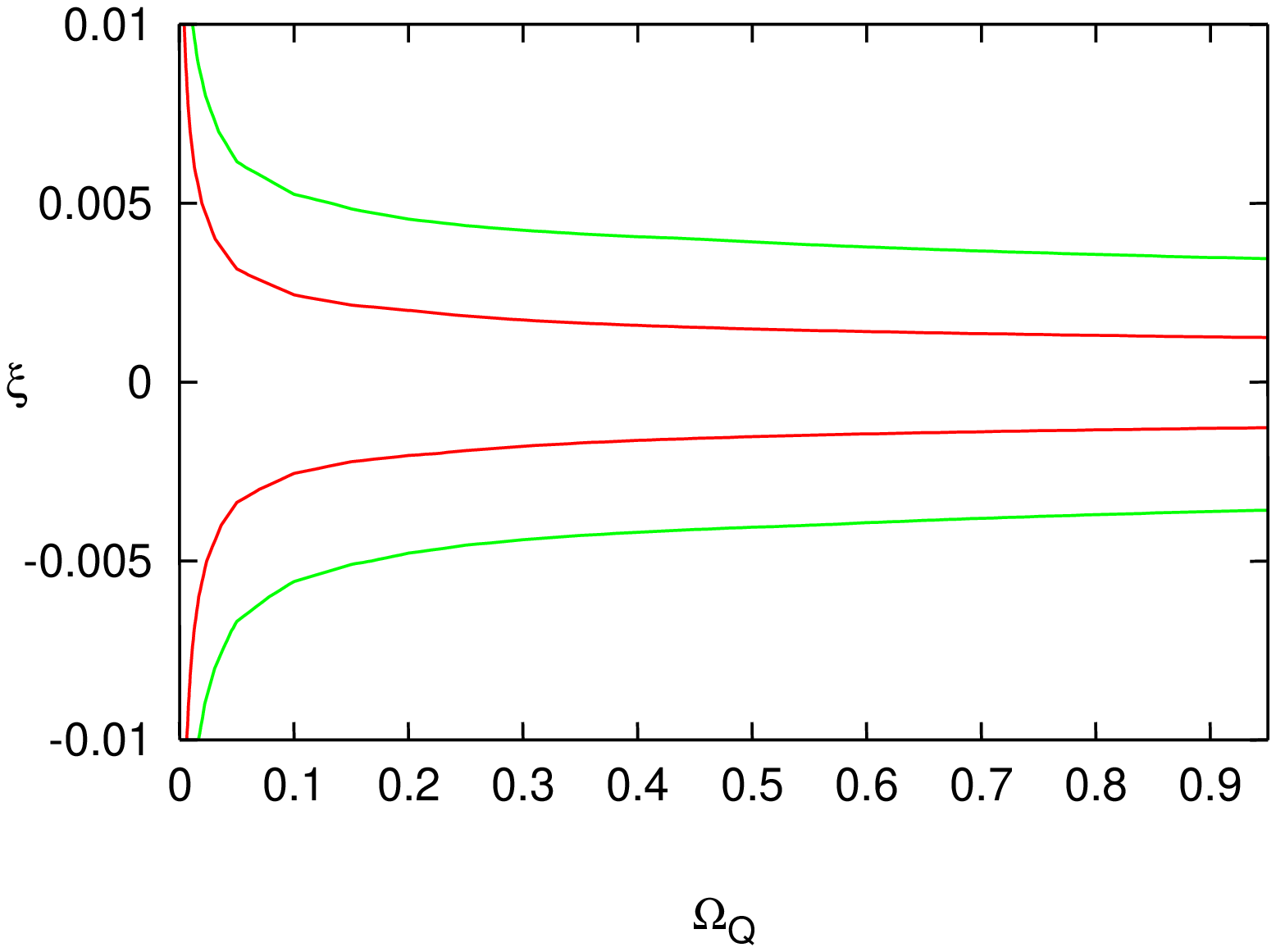,bbllx=135pt,bblly=150pt,
       bburx=500pt,bbury=630pt,angle=270,width=3.5in}}
\caption{Summary of the constraints on the parameters ($\xi$,
  $\Omega_\QUINT^0$) for an inverse power law potential with $\alpha =
  11$ (left) and for a SUGRA potential with $\alpha = 11$ (right). The
  allowed region lies between the lines of same color. The red lines
  represent the constraint arising from the $\gamma_\PN$ post
  Newtonian parameter [see \EQN{\ref{8}}], the green line is from
  $\beta_\PN$, the blue line from the time-variation of
  $G_{\NEWT_\EFF}$, and the purple line from nucleosynthesis. In both
  cases, the most stringent constraint comes from $\gamma_\PN$ and are
  stronger for high $\Omega_\QUINT$.}
\label{fig1}
\end{figure}

In Fig.~\ref{fig1}, we sum up all the preceding constraints in the
plane $(\Omega_\QUINT^0,\xi)$ both for the inverse power law
(\ref{01}) and SUGRA potentials~(\ref{sugra}). Since $\phi \sim M_\PL$
today, $F(\phi_0) - 1 \propto \xi$ and $\xi$ is then a direct
measurement of the deviation from general relativity today from which
it follows that $\xi$ must be very small today (as first pointed out
in Ref.~\cite{chiba99}). The constraints are stronger for inverse
power law potentials. This is easily understood if one notices that
for small $\xi$, the equation of state parameter $\omega$ is higher in
the SUGRA case. This translates into a stronger time dependence of the
quintessence field, which in turn puts more stringent constraints on
the coupling between the scalar field and the metric. This remark
therefore extends the bounds obtained in Ref.~\cite{chiba99}.  In both
cases, the most stringent constraint arises from the constraint on the
post-Newtonian parameter $\gamma_\PN$ and all these constraints become
stronger for high $\Omega_\QUINT$. Another interesting point would be
to study how these bounds depend on $\alpha$. For the Ratra-Peebles
potential, this was already addressed in Chiba's work: the bound on
$\xi$ varies as $\alpha^{-1}$, which can be checked numerically. For
the SUGRA potential, the answer is even simpler: the dynamics of the
quintessence field varies very weakly with
$\alpha$~\cite{brax99}. Therefore, the bound on $\xi$ are not
significantly dependent on $\alpha$.
 
\subsubsection{Exponential coupling}

As a second example, we shall consider the class of scalar-tensor
models in which the coupling function $A$ is given by
\begin{equation}
\label{24}
A_\EX (\widetilde \phi) = B e^{- \beta \widetilde \phi} ,
\end{equation}
with either the inverse power law or SUGRA potentials as defined in
Eqs.~(\ref{01}) and~(\ref{sugra}).

To compare with, the standard Damour-Nordtvedt mechanism of attraction
of scalar-tensor theories towards general relativity, one requires the
coupling constant $\alpha$ to drive the field towards its minimum
where $\phi = 0$.  In the case of quintessence, the field evolves
toward infinity at late time and we would need to consider a coupling
function $\alpha$ such as
\begin{equation}
\label{24b}
\alpha_\DN (\widetilde \phi) = - B e^{- \beta \widetilde \phi} ,
\end{equation}
that tends to zero at infinity in order to converge toward general
relativity.  The function $A$ is easily obtained by integrating
\EQNS{(\ref{coupling})}, (\ref{24b}) to give
\begin{equation}
\ln \frac{A_\DN}{A_{\DN, i}}
 = - \frac{1}{\beta} 
     \left[ \alpha_\DN (\widetilde \phi) - \alpha_\DN (\widetilde \phi_i)
     \right] ,
\end{equation}
where the subscript $i$ refers to some initial time.  On the contrary,
the class of models~(\ref{24}) does not exhibit the double attractor
mechanism~\cite{bartolo} because $\alpha$ and hence, the PN parameter
$\gamma_\PN$ are constant: one has $\alpha_\EX = - \beta$,
$\gamma_\PN^\EX \sim 1 - 2 \beta^2$, and $\beta^\EX_\PN =
0$. therefore, as long as $\beta$ is sufficiently small (we shall take
$\beta = 0.025$ in the following), this model can be compatible with
the Solar system constraints. Note that in this model, the constancy
of $\alpha_\EX$ requires that the coupling function $F$ is a
polynomial of degree $2$, exactly as in the first example.  The class
of models~(\ref{24b}) can indeed be easily studied along similar
lines.

\begin{figure}[ht]
\centerline{
\psfig{file=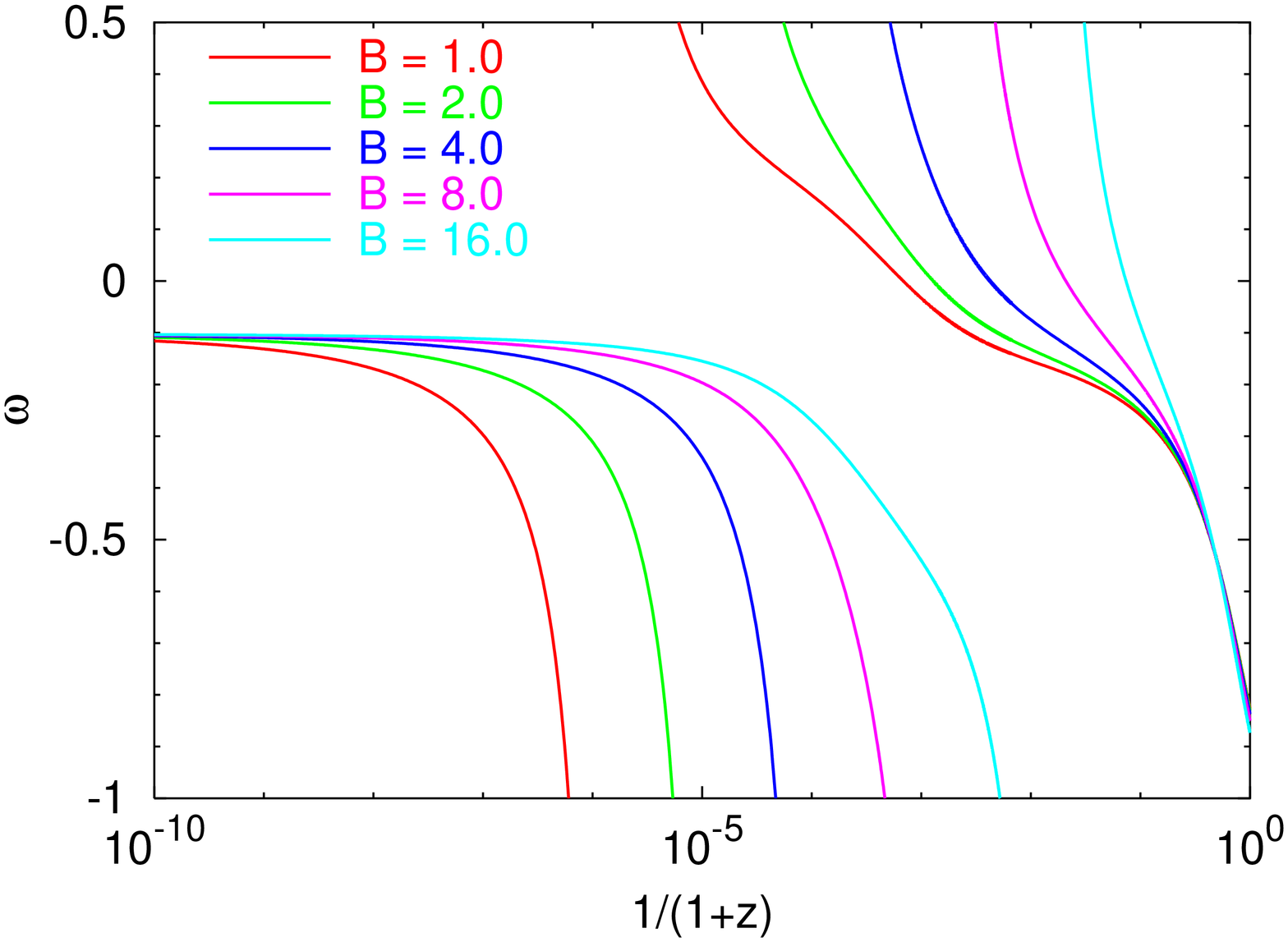,angle=270,width=3.5in}
\psfig{file=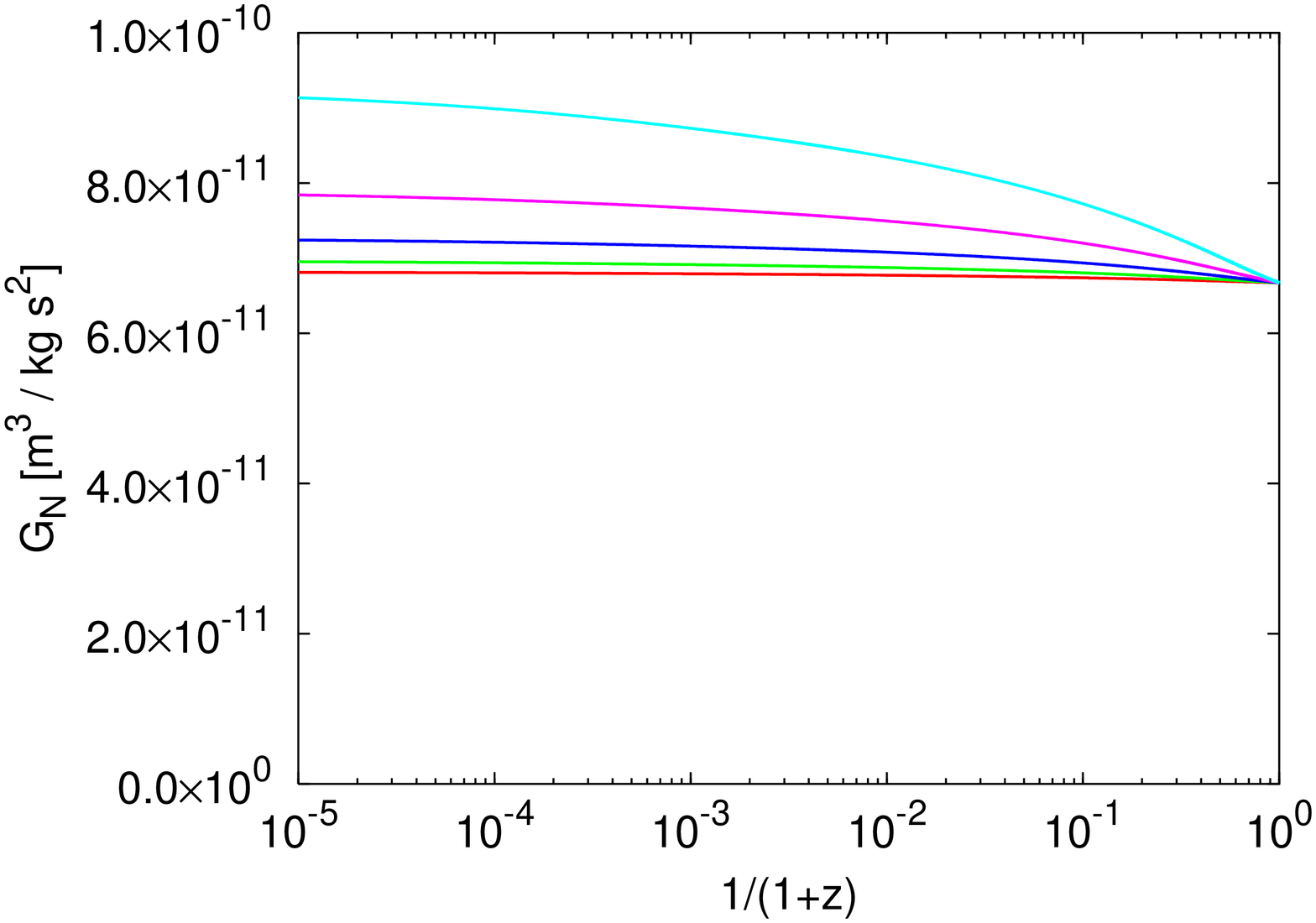,angle=270,width=3.5in}}
\caption{Evolution of the quintessence equation of state parameter
  $\omega_\QUINT$ (left) and of the Newton ``constant''
  $G_{\NEWT_\EFF}$ (right) as a function of the redshift $z$ for
  $\Omega_\QUINT^0 = 0.7$, $\beta = 0.0025$, and various values of the
  coupling parameter $B$.}
\label{figevc}
\end{figure}

\begin{figure}[ht]
\centerline{
\psfig{file=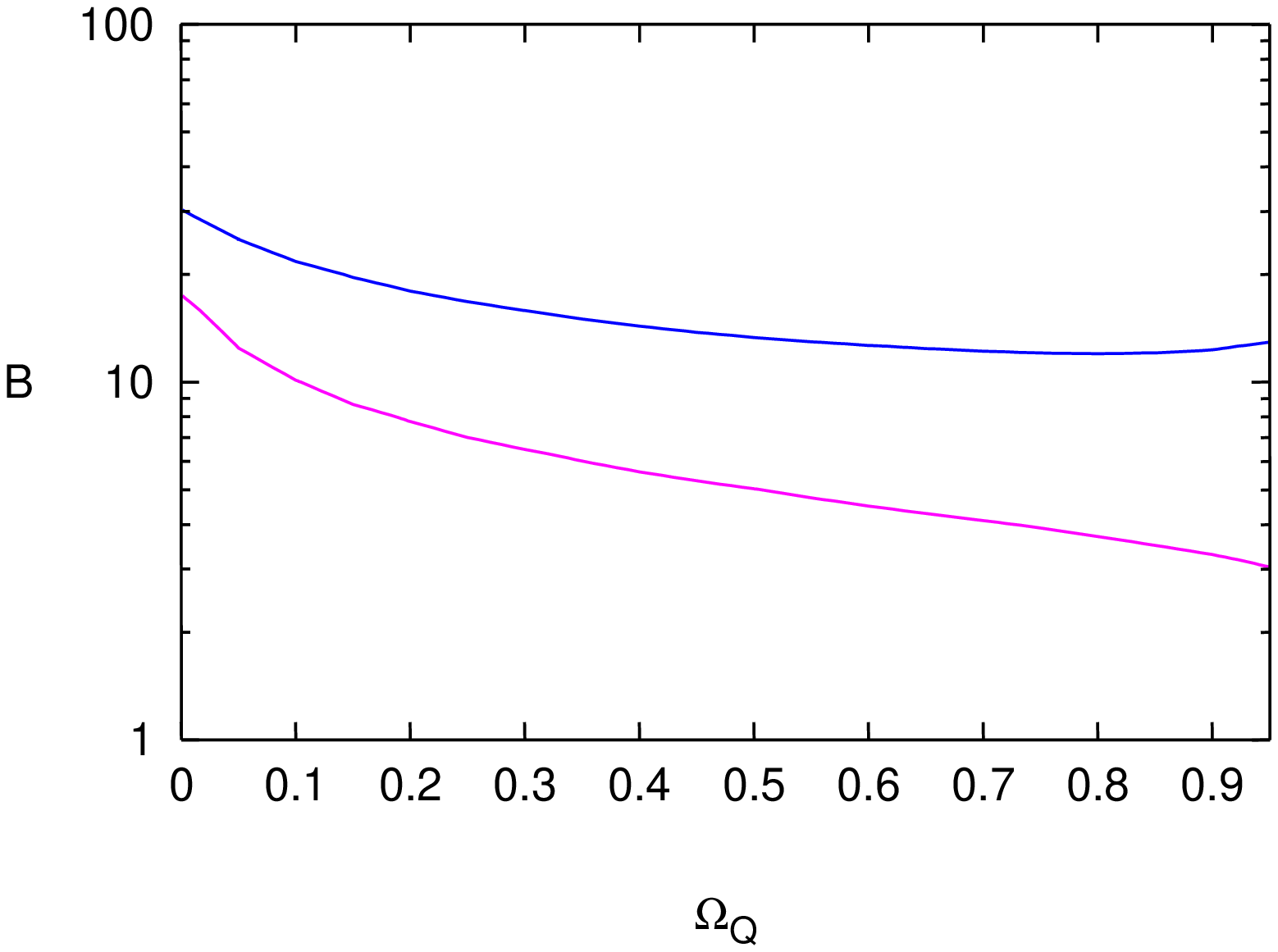,bbllx=135pt,bblly=150pt,
       bburx=500pt,bbury=630pt,angle=270,width=3.5in}
\psfig{file=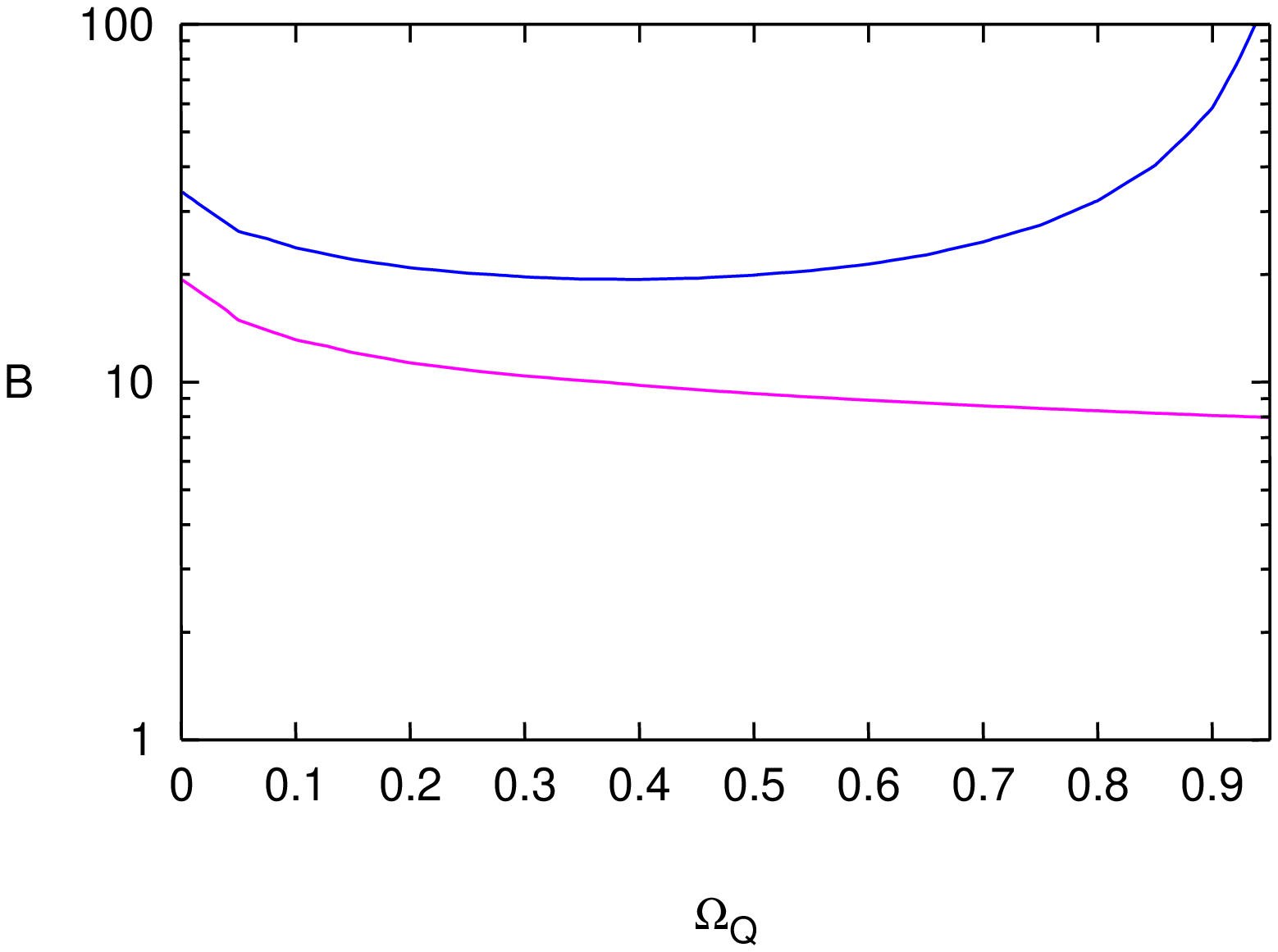,bbllx=135pt,bblly=150pt,
       bburx=500pt,bbury=630pt,angle=270,width=3.5in}}
\caption{Summary of the constraints on the parameters
($B,\Omega_\QUINT^0$) in the case of a Ratra-Peebles (left) and SUGRA
(right) potential with $\alpha = 11$.  The color codes are the same as
for Fig.~\ref{figevc}.}
\label{fig2}
\end{figure}

In Fig.~\ref{figevc}, we show the evolution of the equation of state
parameter $\omega$ and the Newton constant with time. In
Fig.~\ref{fig2}, we sum up all the constraints detailed above in the
plane $(\Omega_\QUINT^0, B)$. As already noted, the constraint on the
PN parameter $\beta_\PN$ is trivially verified, and the constraint on
$\gamma_\PN$ is also satisfied as long as the parameter $\beta$ in
Eq.~(\ref{24}) is small. Therefore, only the two other constraints
play a role, in contrast with the former case.

\subsection{Supernovae data}
\label{IIB}

The use of type Ia supernovae to constraint the cosmological
parameters (and hence the claim that our universe is accelerating)
mostly lies in the fact that we believe that they are standard candles
so that we can reconstruct the luminosity distance vs redshift
relation and compare it with its theoretical value. In a scalar-tensor
theory, we have to address both questions, \IE, the determination of
the luminosity distance vs redshift relation (Sec.~\ref{IIB1}) and the
property of standard candle since two supernovae of different redshift
are feeling a different gravitational coupling constant and may not be
standard candles anymore (Sec.~\ref{IIB2}).

\subsubsection{Luminosity distance in scalar-tensor theories}
\label{IIB1}

To derive the luminosity distance ($\dd_\LUM$) vs redshift ($z$)
relation needed to interpret the supernovae data, we rewrite
Eq.~(\ref{6}) as
\begin{equation}
\label{20}
\left(\frac{\Hconf}{\Hconf_0}\right)^2
 \equiv   E^2(x)
 =        \frac{\kappa_\EFF}{\kappa_\EFF^0}
          \left[  \Omega_\MAT^0 x^{- 1}
                + \Omega_\RAD^0 x^{- 2}
                + \Omega_\QUINT^0 
                  \frac{\rho_\QUINT (x)}{\rho_\QUINT (1)} x^2
          \right]
        + \Omega_\COURB^0 ,
\end{equation}
where $x \equiv 1 / (1 + z)$, and $\Omega_\COURB^0 \equiv - K /
\Hconf_0^2$.  The metric of the constant time hypersurfaces is
decomposed as
\begin{equation}
\gamma_{ij} \dd x^i \dd x^j
 = \dd \chi^2 + \SK{\COURB}^2 (\chi) \dd^2 \omega ,
\end{equation}
where $\dd^2 \omega$ is the infinitesimal solid angle and $\SK{\COURB}
(\chi) \equiv [\sin (\sqrt{K} \chi) / \sqrt{K}, \chi, \sinh (\chi
\sqrt{-K}) / \sqrt{-K}]$ for $K$, respectively, positive, zero or
negative.  With these notations, the luminosity distance is given
by~\cite{peebles93}
\begin{equation}
\label{dlumz}
\dd_\LUM [z; \Omega_0, \Omega_\QUINT^0, \xi]
 = (1+z) \frac{c}{H_0} 
   \SK{-\Omega_\COURB^0} \left[\int_0^z \frac{\dd z'}{(1 + z') E(z')}
                         \right] .
\end{equation}

\subsubsection{SN~Ia in scalar-tensor theories}
\label{IIB2}

The standard lore is to compare this result with type Ia
supernovae data to extract the cosmological parameters assuming
that they are standard candles in the sense that their light curve
does not depend on the supernovae and in particular of $z$. A time
varying effective gravitational constant can affect this picture
at least in two ways~\cite{garcia99} by changing
\begin{enumerate}
  
\item the thermonuclear energy release, since the luminosity at the
  maximum of the light curve is proportional to the mass of nickel
  synthetized,
  
\item the time scale of the supernova explosion and hence the width of
  its light curve.

\end{enumerate}
As we pointed out in the Introduction, it was shown in
Ref.~\cite{gef00b} that the relation~(\ref{20}) for a flat Cold Dark
Matter model with a cosmological constant ($\Lambda$CDM) in the
framework of general relativity can be mimicked by a scalar-tensor
theory with $V = \Lambda = 0$ up to a given redshift. But, this study
did not include the fact that in scalar-tensor theories one, as will
be shown here, can not directly use the luminosity distance vs
redshift relation inferred in the framework of general
relativity. Hence, when comparing any scalar-tensor theory to
supernovae type Ia (SN~Ia) data, one needs a modified magnitude vs
redshift relation taking into account the effects listed above. The
goal of this section is not to explain the recent SN~Ia data by
replacing the cosmological constant by a scalar-tensor theory but
rather to know how to deal with these data in such a framework and to
evaluate the effect of the coupling on the precision of the
determination of the cosmological parameters.

To discuss these issues, we recall a simple ``one zone'' toy
model~\cite{arnett,robm} of an expanding sphere of uniform temperature
$T$ and density and of radius and mass $R_\ENV$ and $M_\ENV$ for the
supernovae light curve which encapsulates the main features of
dependence in the gravitational constant $G$, even if this model is
nothing but a toy model. The observed light curve is obtained from the
nonadiabatic evolution of the thermal energy $E_\THERM = 4 \pi a T^4
R_\ENV^3 / 3$ of the radiation dominated envelope
\begin{equation}
\label{ee}
\dot E_\THERM + 4 \pi P R^2_\ENV \dot R_\ENV = - L + L_* ,
\end{equation}
where $L \simeq E_\THERM / \tau_\DIFF$ is the bolometric luminosity.
$L_* = L_{* \, 0} \exp[- t / \tau_*]$ is the radioactive energy input
(from $\ATOME{}{56}{\NIC}{}{} \rightarrow \ATOME{}{56}{\COB}{}{}
\rightarrow \ATOME{}{56}{\FER}{}{}$, where $\tau_*$ represents the
corresponding nuclear process half life time).  $P = a T^4 / 3$ is the
radiation pressure and the thermal energy $E_\THERM$ and the photon
diffusion time scale $\tau_\DIFF$ is given by
\begin{eqnarray}
\label{21}
\tau_\DIFF
 = 3      \frac{R_\ENV^2}{\lambda_\gamma}
   \simeq 9 \frac{\kappa_\THM M_\ENV}{4 \pi R_\ENV} ,
\end{eqnarray}
where $\kappa_\THM$ is the Thomson opacity and $\lambda_\gamma$ the
photons free mean path. Since the temperature scales as $1/R_\ENV$, it
follows that
\begin{equation}
\label{22}
E_\THERM = \frac{M_0}{2} \frac{R_0}{R_\ENV} ,
\end{equation}
where $R_0$ and $M_0$ are respectively the radius of and the mass of
the progenitor. \EEQN{(\ref{ee})} yields the equation for the
luminosity $L$
\begin{equation}
\label{ff}
\frac{R_0}{R_\ENV} \dot L = (- L + L_*) \frac{1}{\tau_\DIFF^0} ,
\end{equation}
where $\tau_\DIFF^0$ is the initial diffusion time scale when $R_\ENV
= R_0$.  If the envelope expands at a constant velocity $v_\EXP
\equiv \dot R_\ENV$, an analytic solution to Eq.~(\ref{ff}) was found by
Arnett~\cite{arnett} as
\begin{equation}
L = L_\PEAK e^{- u (x)} + L_*^0 \Omega(x, y, w) , 
\end{equation}
where, from Eqs.~(\ref{21}) and~(\ref{22}), $L_\PEAK \simeq E_\THERM /
\tau_\DIFF \simeq 2 \pi R_0 / 9 \kappa_\THM$ is the luminosity at the
maximum of the light curve, $u (x) \equiv w x + x^2$, $x \equiv t /
\tau$, $y \equiv \tau / 2 \tau_*$, $\omega \equiv \tau / \tau_\DIFF^0$,
and $\tau$ is the characteristic time of the SN given by
\begin{equation}
\tau^2 = 2 \frac{R_0}{v_\EXP} \tau_\DIFF^0 ,
\end{equation}
and the function $\Omega$ takes the form
\begin{equation}
\Omega (x, y , w)
 \equiv e^{- u (x)} \int_0^x (w + 2 z) e^{- 2 z y + u (z)} \dd z .
\end{equation}
The expansion velocity can be obtained via the conservation of
energy as 
\begin{equation}
\label{consistency_v}
v_\EXP^2 = 2 \frac{M_\SS{\NIC}}{M_0} .
\end{equation}
In order the model to be consistent, we are therefore obliged to
assume that $M_\SS{\NIC} \leq \frac{1}{2} M_0$.  From the above
equation, it follows that $\tau$ behaves as
\begin{equation}
\tau^2
 \simeq \frac{9 \kappa_\THM}{2 \pi \sqrt{2}} 
        \left(\frac{M_0^3}{M_\SS{\NIC}}\right)^{1 / 2} .
\end{equation}
Assuming that the progenitor has the Chandrasekar mass $M_\CHAND$,
$R_0$ is the Chandrasekar radius, and assuming that the mass of nickel
scales as $M_0$, we deduce that
\begin{equation}\label{38}
L_\MAX = L_\MAX^\GR \left(\frac{G_\EFF}{G}\right)^{- 3 / 2}
, \quad
\tau = \tau^\GR \left(\frac{G_\EFF}{G}\right)^{- 3 / 4} ,
\end{equation}
since the total energy release is proportional to mass of nickel
formed which is assumed to scale as the progenitor mass and thus as
$G^{- 3 / 2}$.

Such a toy model can, at that stage, describe both SN~I and SN~II.
But, for SNIa, the progenitor is a white dwarf and it follows that
$R_0 \sim 5 \times 10^3 \UUNIT{km}{}$. Then, since
\begin{equation}
\tau_\DIFF^0
 \simeq 10^8 \frac{M_\ENV}{M_\odot}
        \left(\frac{R_0}{5\times 10^3 \UUNIT{km}{}}\right)^{-1}
\end{equation}
and 
\begin{equation}
\tau
 \simeq 71 \frac{M_\ENV}{M_\odot}^{1 / 2}
        \left(\frac{v_\EXP}{10^4 \UUNIT{km}{} \UUNIT{s}{-1}} 
        \right)^{- 1 / 2} ,
\end{equation}
we can conclude that $L_\PEAK \simeq 0$ and that $\omega \sim 10^{-
  7}$. We then choose the typical~\cite{robm} value of the parameters
for SN~Ia to be
\begin{equation}
L_\PEAK = 0
,\quad 
w = 0 .
\end{equation}
It follows that the luminosity curve is well approximated by $L =
L_*^0 \Omega (x, y ,0)$ which can be integrated analytically to give
\begin{equation}
L = L_*^0
    \left(  e^{- 2 x y}
          - e^{x^2}
          + i \sqrt{\pi} y e^{- (x^2 + y^2)}
            \left[\hbox{Erf} (i x - i y) + \hbox{Erf} (i y)\right]
    \right) ,
\end{equation}
where $\hbox{Erf}$ is the error function and where the nuclear rate
are given by
\begin{eqnarray}
\mbox{cobalt: } 
L_* = 1.5 \times 10^{43} M_\SS{\NIC} \UUNIT{erg}{} \UUNIT{s}{-1} 
& ,\quad &
\tau_* = 111 \UUNIT{days}{} ,
\\ 
\mbox{nickel: }  
L_* = 8 \times 10^{43} M_\SS{\NIC} \UUNIT{erg}{} \UUNIT{s}{-1} 
& ,\quad &
\tau_* = 10.1 \UUNIT{days}{} .
\end{eqnarray}

\begin{figure}[ht]
  \centerline{ \psfig{file=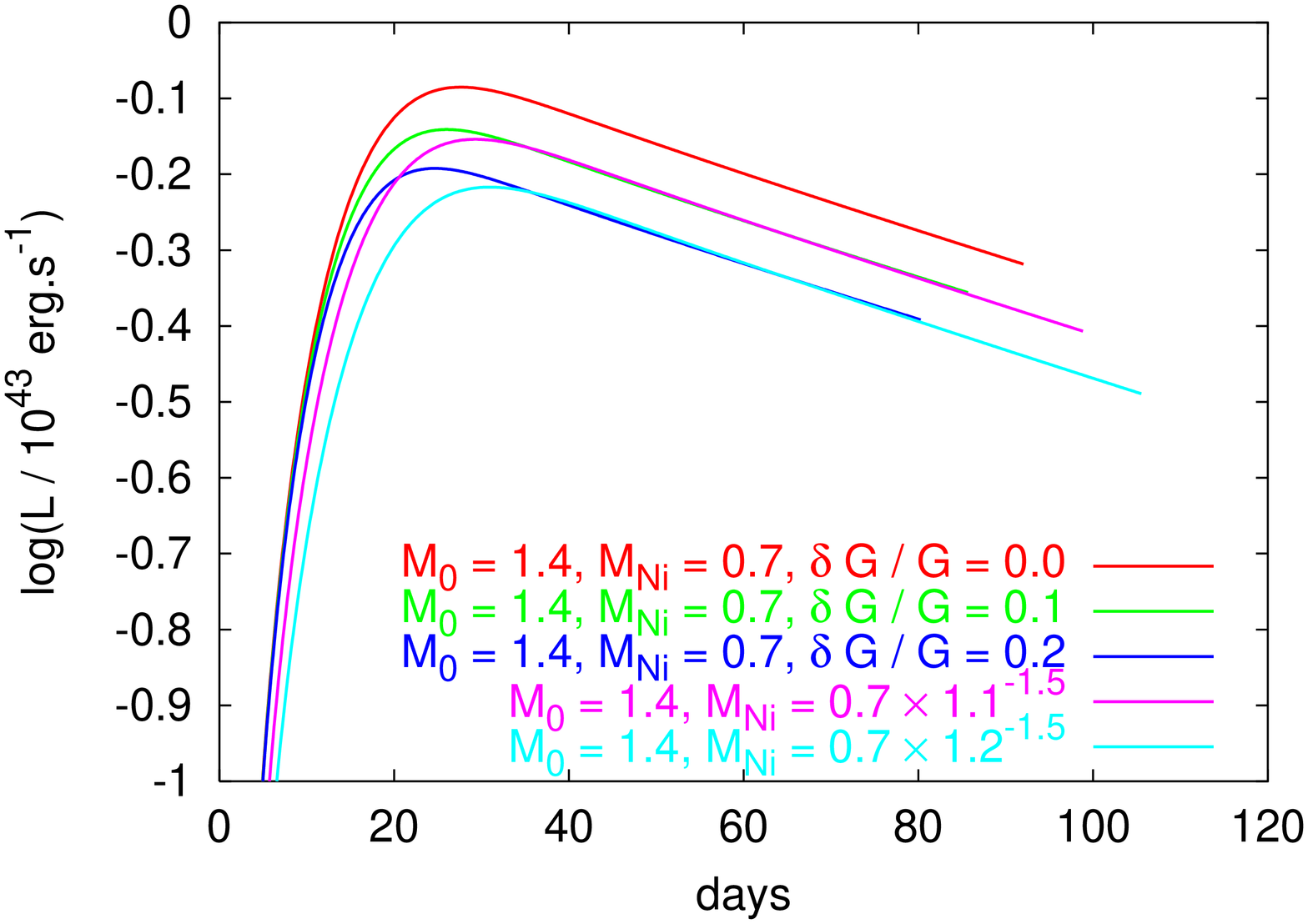,angle=270,width=3.5in}
               \psfig{file=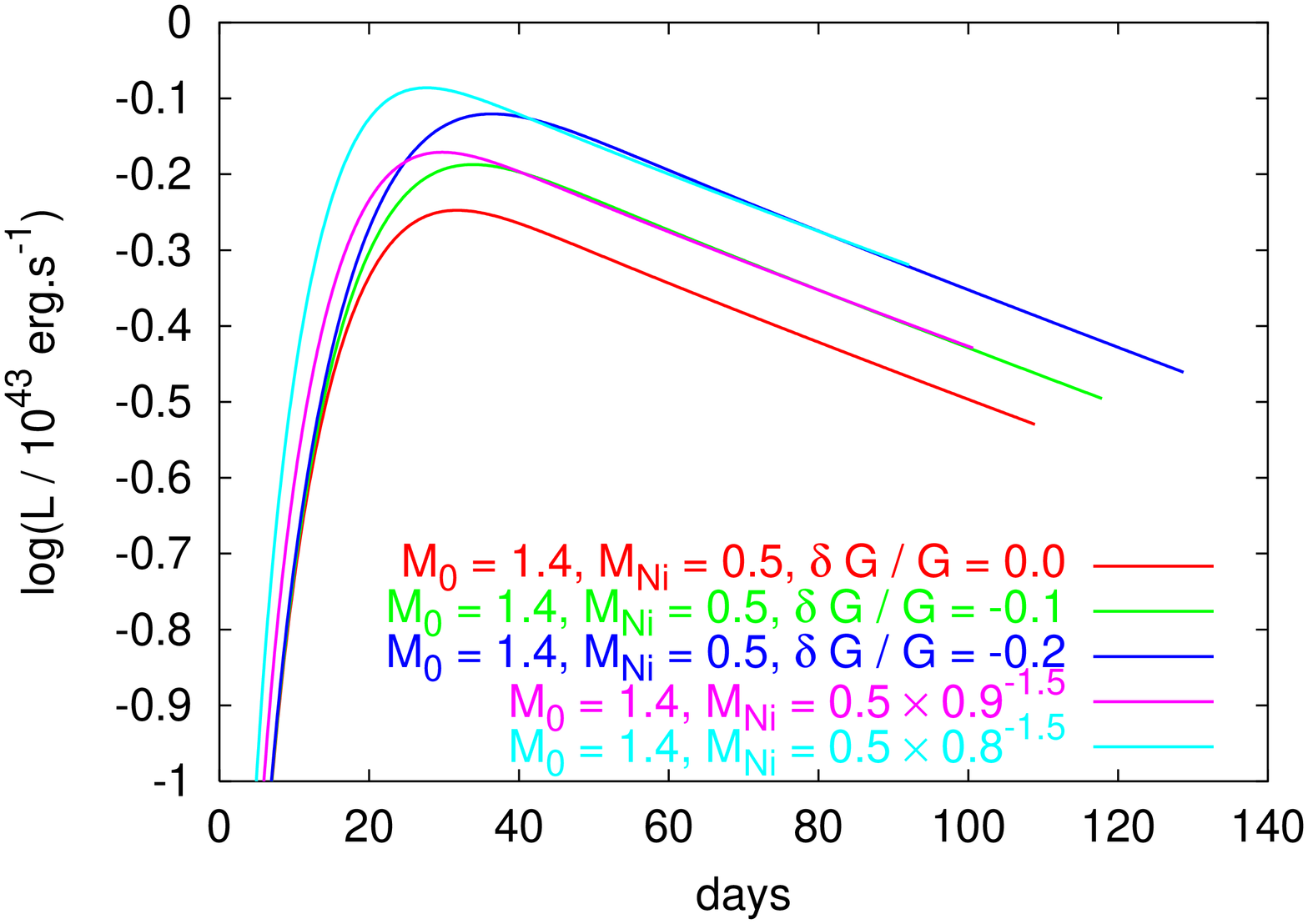,angle=270,width=3.5in}}
\caption{Variation of the light curve of SN~Ia with respect to an
  increase of the gravitational constant in the simple ``one zone''
  toy model presented in this article. A stronger gravitational
  constant accounts for a fainter and narrower light curve (see
  \EQN{(\ref{38})}). The rising part is mainly sensitive to the
  progenitor mass whereas the decaying part is mainly dictated by the
  mass of nickel. Hence each light curve computed with a higher Newton
  constant tends asymptotically toward the light curve of a supernovae
  with smaller nickel mass but with standard value of the Newton
  constant (left), and vice versa (right). [Note that in this last
  case, we have to ensure that the total nickel mass does not exceed
  half of the total mass in order for Eq.~(\ref{consistency_v}) to be
  valid.]}
\label{fig3}
\end{figure}

In Fig.~\ref{fig3}, we depict a standard light curve obtained with
this one zone model and the light curves when $G$ is increased
respectively by $10\%$ and $20\%$. This light curves are compared to
the ones obtained by a variation of the nickel mass synthetised. The
decaying branch is mainly sensitive to the mass of nickel so that an
increase of the gravitational constant implies that the light curve
has a lower maximum and then tends asymptotically towards the light
curve of a supernovae with lower nickel mass.  As a conclusion, if
$G_\EFF$ is $10\%$ larger than $G$ all other parameters being
unchanged, then the luminosity at maximum will be slower by about
$15\%$ and the time scale $\tau$ will be smaller by $7.5\%$ and the
light curve will be narrower. As was realized, supernovae are not
exactly standard candles but, thanks to the correlation between the
time scale of the light curve and the peak luminosity (larger curves
are brighter while narrower are fainter), the dispersion of
$0.15$--$0.2 \UUNIT{mag}{}$ can be corrected by the use of a stretch
factor~\cite{sndata}.

A variation of $G$ affects both the amplitude of the peak and the time
scale and, for instance, makes it narrower and fainter if $G$ grows in
the past. Again, they can be calibrated to extract the luminosity
distance since, once the model is specified, the dependence of the
correction due to the variation of $G$ is known.  The magnitude vs
redshift relation then takes the form
\begin{equation}\label{mag}
m(z) =   {\cal M}_0
       + 5 \log \dd_\LUM [z; \Omega_0, \Omega_\QUINT^0, \xi] 
       + \frac{15}{4} \log \frac{G_\EFF}{G} ,
\end{equation}
if we just take into account the effect on the peak luminosity
(assuming that a stretch factor has been applied yet).  Our result is
compatible with the relation obtained in the case of a Brans-Dicke
theory~\cite{garcia99} for which $G = (4 + 2 \omega) / (3 + 2 \omega)
\phi^{- 1}$. Such a correction was also argued in Ref.~\cite{amendola}
where it was assumed that the peak luminosity scales as $G_\EFF^{-
\gamma}$ and that $G_\EFF = G (1 + t H_0)^m$ leading to what was
referred to as a ``$G$ correction'' in the magnitude vs redshift
relation of the form $\Delta m_G = 2.5 m \gamma \log [1 + \tau(z)]$,
$\tau(z)$ being the look-back time. Under these hypothesis these
authors showed that an increase in $G$ in the past can reconcile the
SN~Ia data with an open, $\Lambda = 0$ cosmology. But, what was not
shown is that such a variation can be cast into a scalar-tensor
framework while respecting all the constraints described in
Sec.~\ref{IIa} and particularly respecting the positivity of the
energy of the graviton~\cite{gef00b}. The phenomenological exponent
$\gamma$ is not a free parameter and has to be determined from the
theory. Here, we claim, on the basis of our toy model, that
$\gamma=3/2$.

\EEQN{(\ref{mag})} gives the general magnitude vs redshift relation in
scalar-tensor theory and has, to be self-consistent, to be used when
comparing an extended quintessence model to the data. Indeed, the one
zone toy model presented above may be thought to be very rough but is
qualitatively correct~\cite{robm}. We have to stress that we neglected
the effect of the scalar field in the SN~Ia dynamics and assumed that
its only effect (or at least dominant) was to change the value of the
gravitational constant. We also point out that we used $G_\EFF$
instead of $G_{\NEWT_\EFF}$ in Eq.(\ref{mag}) but, as shown in
Ref.~\cite{boisseau} they may not differ from more than $10\%$, which
is a $1\%$ effect in our magnitude vs redshift relation.

\subsubsection{Application to two examples}

\begin{figure}[ht]
  \centerline{\psfig{file=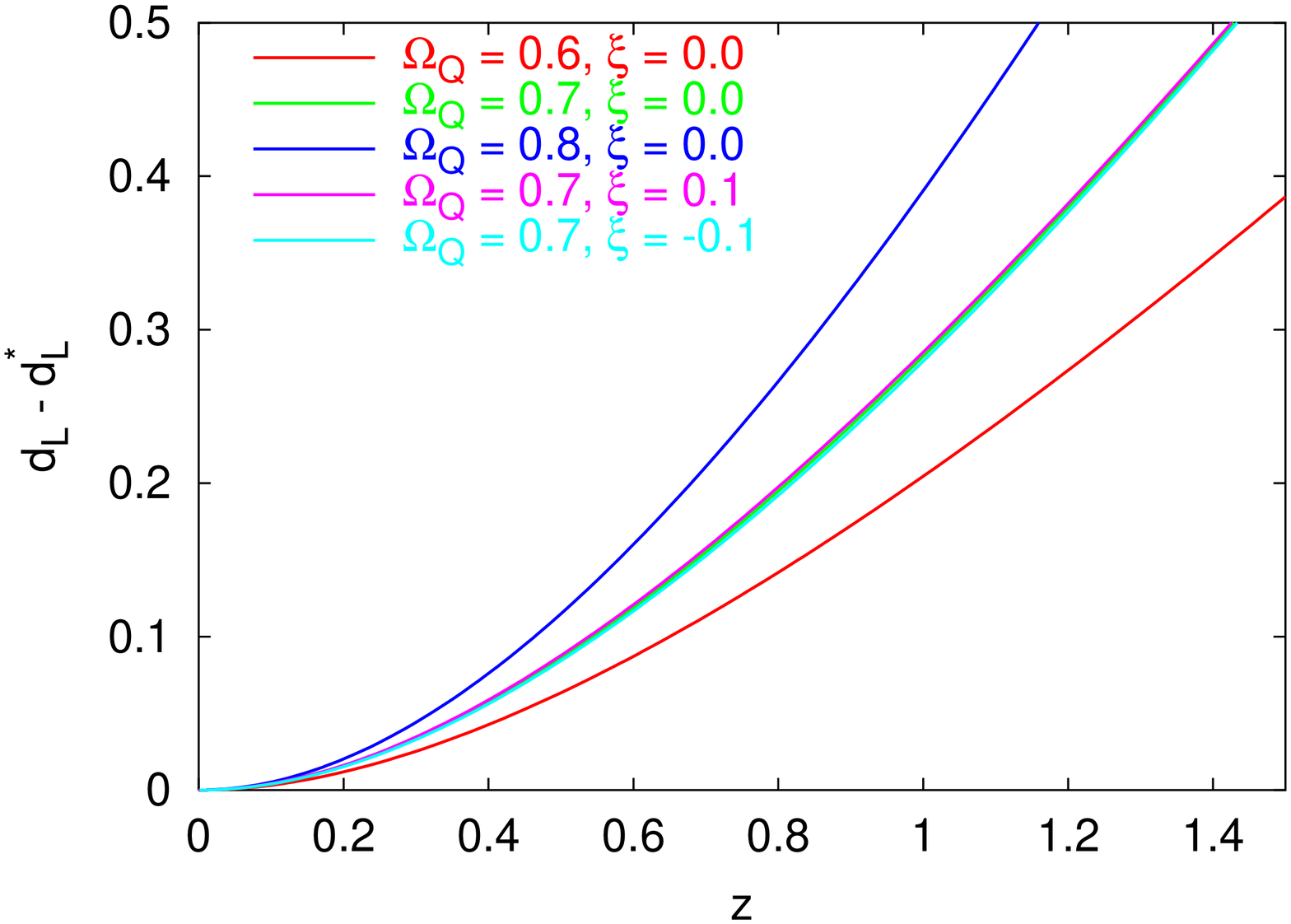,angle=270,width=3.5in}
              \psfig{file=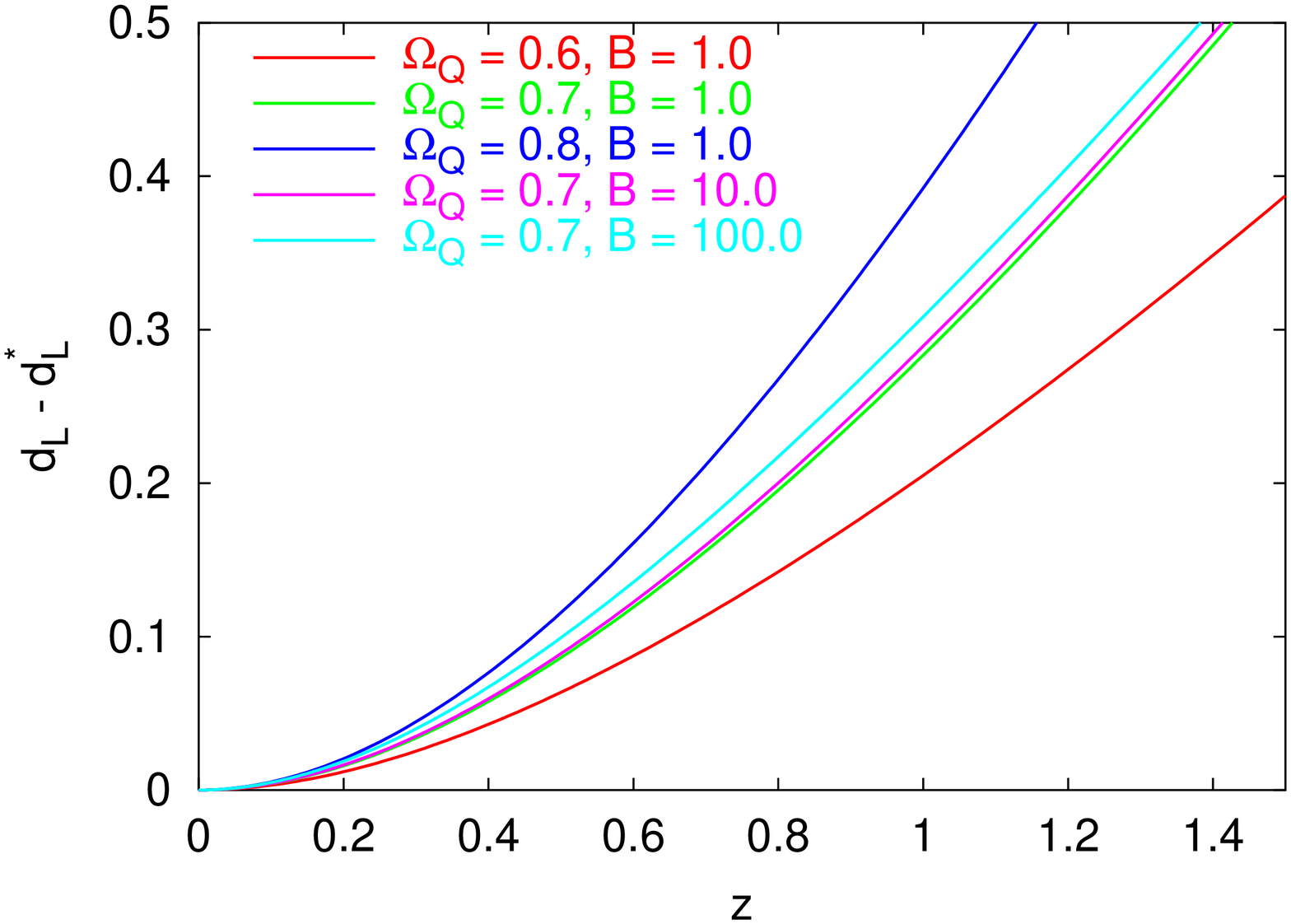,angle=270,width=3.5in}}
\caption{Luminosity distance vs redshift relation for different value of
  $\xi$ (left) and $B$ (right). Both are compared to the luminosity
  distance vs redshift relation $\dd_\LUM^*$ for a matter dominated,
  flat universe. As noted previously, the deviation is more important
  for positive $\xi$ than for negative $\xi$.}
\label{fig4}
\end{figure}

In Fig.~\ref{fig4}, we show the luminosity distance vs redshift
relation for the two models presented above under the assumption that
the universe is spatially flat (\IE, $\Omega_\MAT + \Omega_\QUINT =
1$).  In the case of the nonminimally coupled quintessence field, it
can be noted that the deviation is more important for positive $\xi$
than for negative $\xi$. Note that in any case, the net effect of the
coupling of the luminosity distance is small (the most extreme values
of $\xi$ and $B$ plotted here are already ruled out). However, when
fitting the supernovae data, one must also take into account the
change in the SN absolute luminosity, which can be as large as $30\%$
for the largest variation of $G$ allowed by nucleosynthesis.  We also
note that the distance vs redshift relation is more sensitive to the
coupling $\xi$ for smaller values of the exponent $\alpha$.  From the
Friedmann equation~(\ref{6}) and Eq.~(\ref{4}), we have
\begin{equation}
\dd \eta \propto F .
\end{equation}
Moreover, {\it for constant $\phi$}, increasing $\xi$ increases $F$,
and therefore the distance traveled by a photon is increased. This
explains why the quantity $d_\LUM - d_\LUM^*$ increases with $\xi$ in
Fig.~\ref{fig4}. Unfortunately, the same reasoning does not apply for
the exponential coupling because the hypothesis that the value of the
quintessence field $\phi$ does not significantly vary at fixed
redshift when one varies the parameter $B$ is no longer valid. (Should
the field be at rest today, then the hypothesis would still be valid,
but this is not the case, as shown in Figs.~\ref{fig0}
and~\ref{figevc}.)

\begin{figure}[ht]
  \centerline{\psfig{file=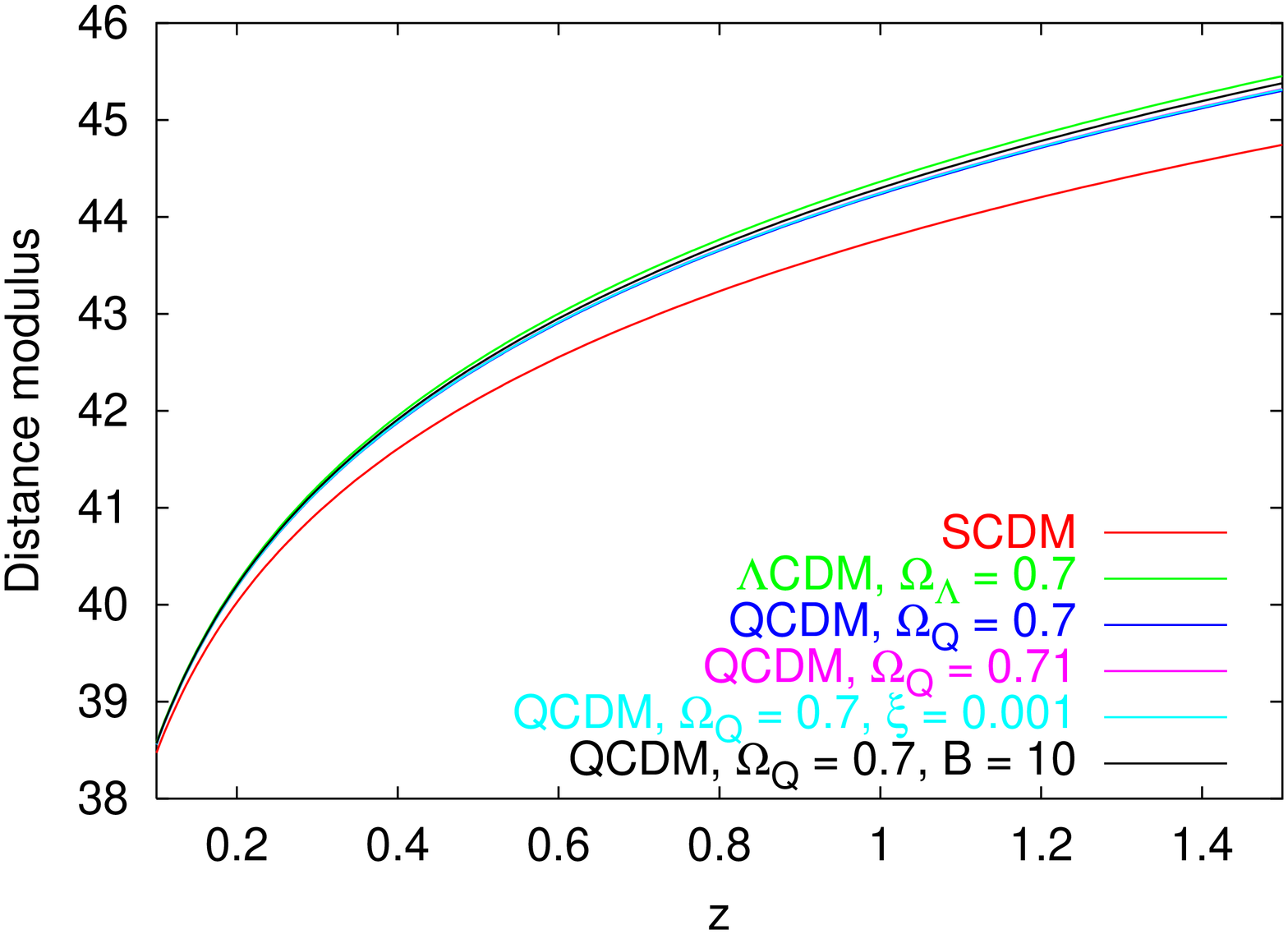,angle=270,width=3.5in}
              \psfig{file=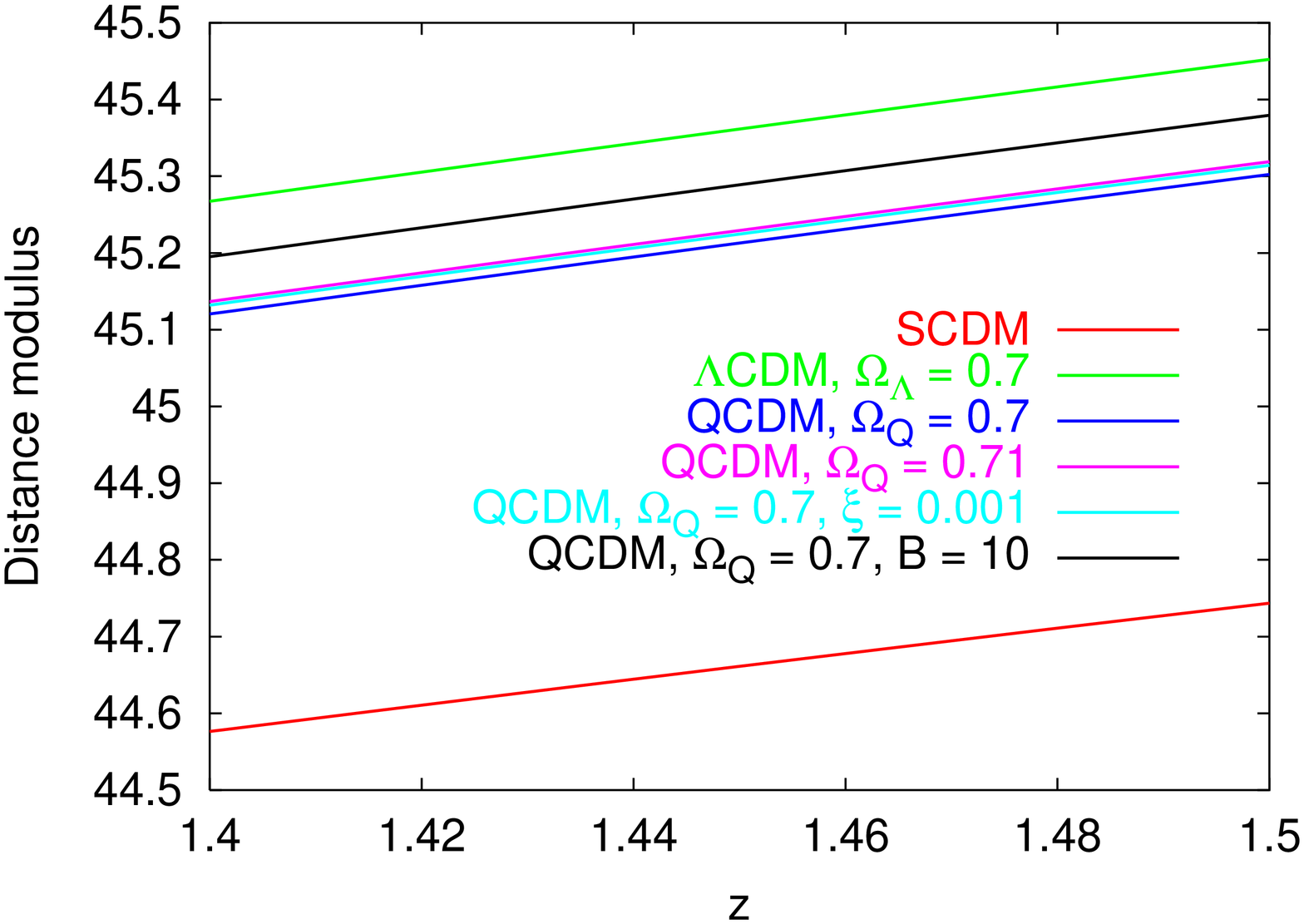,angle=270,width=3.5in}}
\caption{Distance modulus vs redshift relation for different models.
  A comparison between a Standard Cold Dark Matter (SCDM) model, a
  $\Lambda$CDM model, and a Quintessence Cold Dark Matter (QCDM) model
  are shown. Also shown are the maximum allowed deviation from a
  standard QCDM model when considering the two models described in the
  text. The exponential model allows for large deviations (roughly
  equivalent to those obtained by adding $\delta \Omega_\QUINT \sim
  0.05$ in a minimally coupled QCDM model), whereas the non minimal
  coupling is much more constrained. The right panel represents a zoom
  of the left panel for high redshift.}
\label{fig4bis}
\end{figure}

In Fig.~\ref{fig4bis} we present the distance modulus vs redshift
relation taking into account the supernovae luminosity correction due
to a variation of the Newton constant. We compare the amplitude of
this correction to the standard $\Lambda$CDM model. It appears that
the allowed values for $\xi$ do not lead to a significant modification
of the distance modulus. This is related to the fact that this model
is very strongly constrained by the solar system data. On the
contrary, the exponential coupling model allows for a larger deviation
of the distance modulus, mainly because it is not limited by the post
Newtonian constraints. More generally, any model tuned so to evade the
solar system constraints can lead to large deviation in the magnitude
vs redshift relation.

\subsection{Angular distance}
\label{IIC}

Another important observational quantity is the angular distance,
$\dd_\ANG$, which, with the notation of the previous paragraph, is
given by~\cite{peebles93}
\begin{equation}
\dd_\ANG = \frac{\dd_\LUM}{(1 + z)^2} .
\end{equation}
This distance relates the size of an object to the angle under which
it is observed. Naively, the variation of $\dd_\ANG$ can be related to
the position of the first cosmic microwave background acoustic peak.
Note also that $\dd_\ANG (z)$ also enters the probability of
lensing~\cite{gldata}.

\begin{figure}[ht]
\centerline{\psfig{file=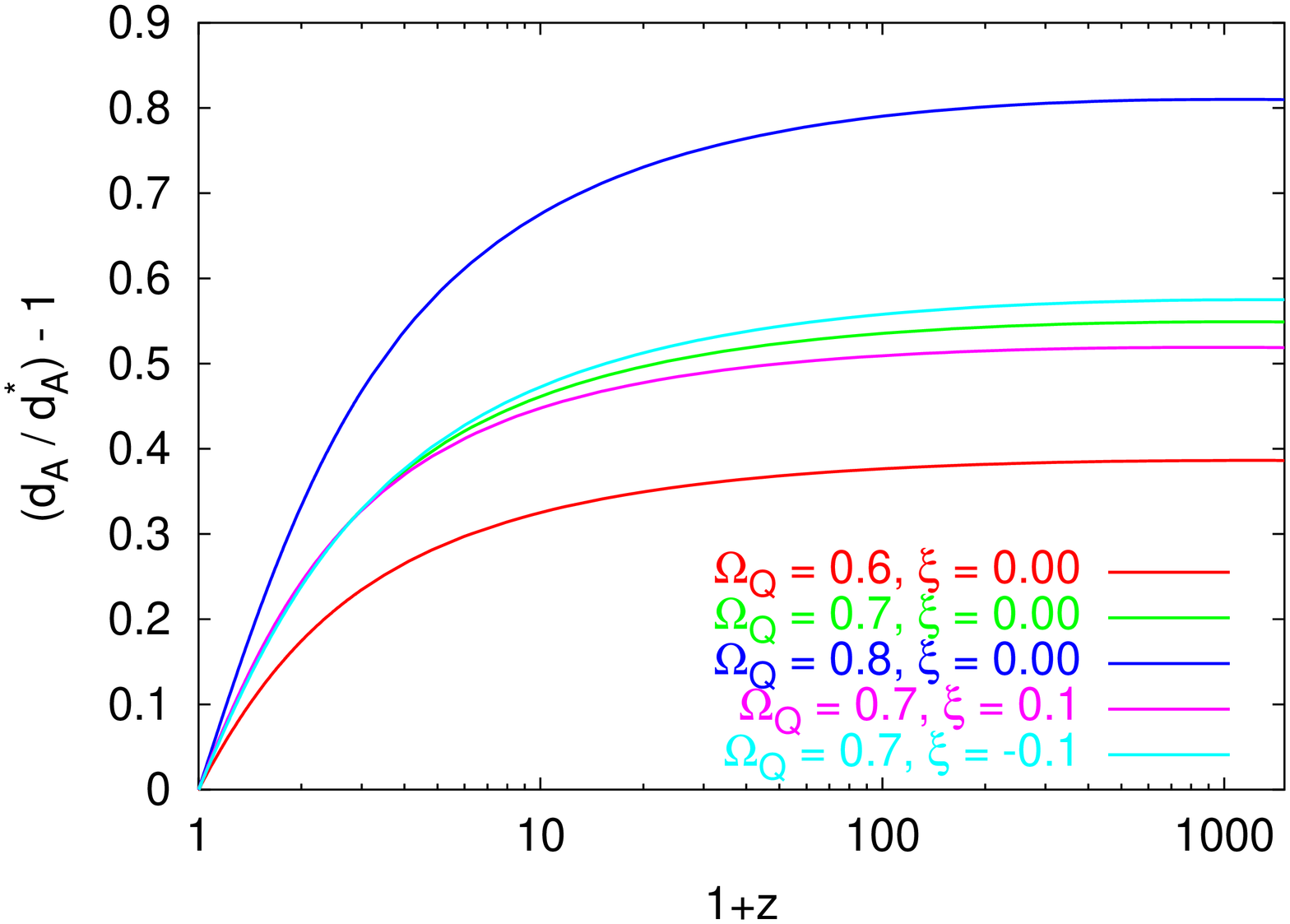,angle=270,width=3.5in}
            \psfig{file=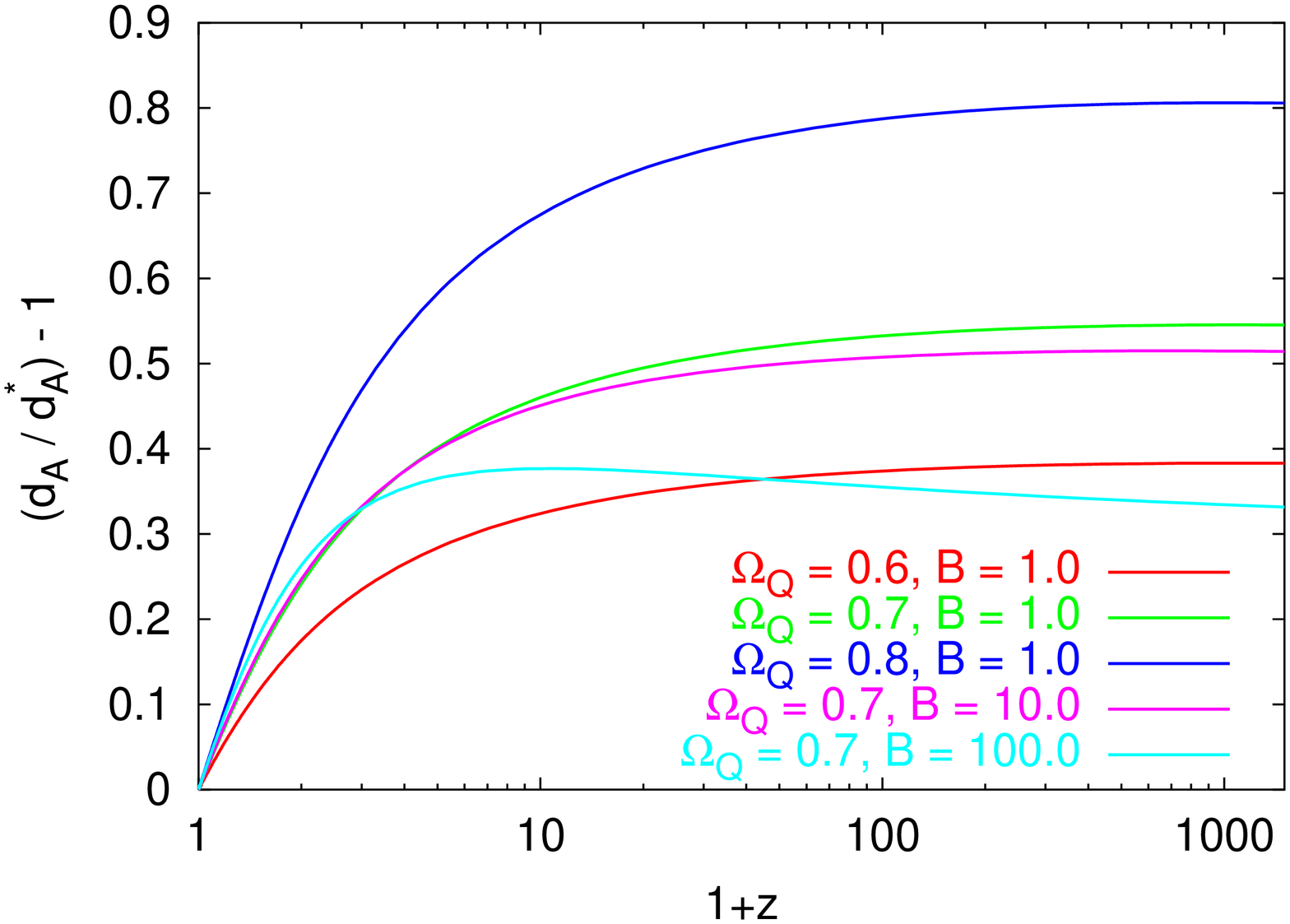,angle=270,width=3.5in}}
\caption{Angular distance vs redshift relation for different values of
  $\xi$ (left) or $B$ (right) compared to $d_\ANG^*$, the angular
  distance vs redshift relation for a matter dominated, flat universe.
  The relative shift in the position of the acoustic peaks is simply
  given by the asymptotic part of the curves. We assumed that
  $\Omega_\QUINT = 0.7$ and used the SUGRA potential with $\alpha =
  11$.}\label{fig6}
\end{figure}

In Fig.~\ref{fig6}, we depict the variation of $\dd_\ANG (z)$ both in
the nonminimally coupled scalar field models and in the exponential
coupling case. Indeed, a complete study of the CMB anisotropies will
be presented in the following section and these results are just a
hint of how the position of the first acoustic peak will be affected
in these models. The relative shift in the position of the acoustic
peaks is simply given by the asymptotic part of the curves of
Fig.~\ref{fig6}. As for the luminosity distance, the effect of the
coupling is rather small. However, the variation of $G$ will lead to
an important modification of the peak position as we shall see later.

\section{Properties of the perturbations}
\label{IV}

The background and perturbation equations in a scalar-tensor
quintessence model are presented in the appendices.  Not that these
equations are not restricted to the case of the nonminimally coupled
scalar field and can be extended easily to the exponential coupling
example by simply setting
\begin{equation}
\xi f = \frac{F(\phi) - 1}{2 \kappa} ,
\end{equation}
since, as emphasized in the Introduction the splitting~(\ref{3}) is
completely general as long as the function $f$ has not be chosen.
Technically, one obtains a set of equations slightly more complicated
than in the minimally coupled case, but which can still easily be
solved numerically. We shall first discuss the main properties of the
scalar field perturbations before turning to cosmic microwave
background properties in the next section.

As for the minimally coupled case, one of the main concern is about
the choice of initial conditions for the perturbations. It happens
that this problem is not relevant since the long wavelength scalar
field perturbations follow an attractor for the following reason.
First, let us consider the unperturbed and perturbed Klein-Gordon
equations
\begin{eqnarray}
\ddot \phi + 2 \Hconf \dot \phi + a^2 V' & = & - \xi a^2 R f' , \\
   \ddot {\delta \phi} - \Delta \delta \phi
 + 2 \Hconf \dot {\delta \phi} + a^2 V'' \delta \phi - 2 \Chi \ddot \phi
 + \dot \phi \left[ \Delta \frac{\Phi}{\Hconf} - \dot \Chi - 4 \Hconf \Chi 
             \right]
 & = & - \xi (a^2 R f'' \delta \phi + a^2 \delta R f' ) ,
\end{eqnarray}
where $\Phi$ is one the two Bardeen potential and $\Chi$ is a
combination of the two Bardeen potentials (see the appendices for more
details).  For an inverse power law potential, one can show that there
exists some particular solution to the unperturbed Klein-Gordon
equation, the so-called tracking solution~\cite{ratra}, for which the
field evolves according to some power law (the exponent of which
depends on the exponent of the potential). Then, the stability of this
particular solution --- which is the most useful feature of the
quintessence scenarios --- is determined by the properties of the
``perturbed'' expression of the Klein-Gordon equation. This
``perturbed'' equation actually describes the small departures of an
homogeneous field from its tracking solution, and reads
\begin{equation}
   \ddot {\delta \phi}
 + 2 \Hconf \dot {\delta \phi}
 + a^2 V'' \delta \phi 
 =  - \xi a^2 R f'' \delta \phi .
\end{equation}
Obviously, this equation is extremely similar to the above ``real''
perturbed Klein-Gordon equation: we simply have neglected the metric
perturbation terms, and considered the large wavelength limit $k \to
0$. The very nice feature of inverse power law potential is that the
solutions to this equation tend to 0. Therefore, the solutions of the
homogeneous part of the real perturbed Klein-Gordon equation in the
large wavelength limit tend to 0.  Now, when one takes into account
the metric perturbations, remembering the fact that they are constant
in the long wavelength limit (see, \EG, Ref.~\cite{bunn}), it appears
that the relevant quantities, such as $\delta_\MC$ (the density
contrast of the quintessence fluid) will tend toward constants which
will simply be linear combinations of the metric perturbations.
Although not explicitly stated, this is what was shown in
Ref.~\cite{brax00}.  Finally, in the short wavelength limit, the
Laplacian term ensure that the field follows a wave equation, which is
actually damped by the expansion.

\begin{figure}[ht]
  \centerline{ \psfig{file=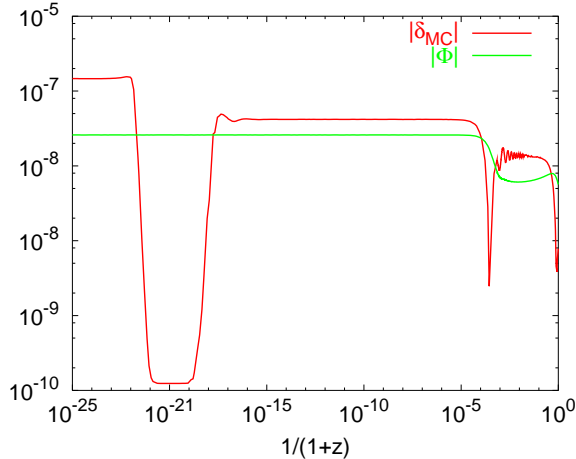,angle=270,width=3.5in}}
  \caption{Evolution of the scalar field density contrast as a
  function of redshift for a non minimally quintessence model (SUGRA
  potential with $\alpha = 11$, power law coupling, and $\xi =
  0.06$). We have represented the evolution of the Fourier mode of the
  quantity $\delta^\flat_\MC (k, z)$ for the wave number $k = 0.1 h
  \UUNIT{Mpc}{-1}$.  By comparison, we have also shown the Bardeen
  potential $\Phi$.  The quintessence field reaches the attractor
  around $z \sim 10^{17}$. After this epoch, the quintessence density
  contrast reaches a value depending only of the potentials $\Phi$ and
  $\Psi$. It subsequently evolves only when the mode has entered into
  the Hubble radius, where it experiences damped oscillations (after
  $z \sim 10^5$) around another value also depending on $\Phi$ and
  $\Psi$. The quintessence density contrast can be very large before
  having reached the attractor.  This is, however, not a problem,
  because the quintessence density contrast $\Omega_\MC$ is small at
  this epoch.  Note that the Bardeen potential has a nontrivial
  behavior at late times both because of the domination of the
  quintessence field (which explains the decay after $z \sim 3$) and
  because of the variation of the Newton constant. }
\label{fig_last}
\end{figure}

As a conclusion, as long as the field is subdominant (which is the
case when one fixes the initial conditions and in the tracking
regimes), its perturbations follow an attractor, thus solving the
problem of the choice of the initial conditions, and the field does
not show any unstable behavior, as illustrated in Fig.~\ref{fig_last}.


\section{Cosmic microwave background and matter power spectrum}
\label{V}

Appendix~\ref{C} gives the derivation of the modified equations for
the evolution of the cosmological perturbations. As explained in this
appendix, the two most important modifications induced by the
quintessence field are (i) a modification of the background equations
which lead to the domination of the scalar field today and (ii) a
subsequent modification of the late time evolution of the
gravitational potentials.  We have implemented the scalar-tensor
quintessence equations in a Boltzmann code (which uses the
line-of-sight integration method, see, \EG,
Refs.~\cite{cmbfast,huwhite}) and computed matter power spectra as
well as CMB anisotropies.  We shall now discuss the observable
consequences of these models.

Two effects of a quintessence (minimally coupled or not) have already
widely been discussed in the literature~\cite{brax00,quintCMB}. These
are the following:
\begin{itemize}
  
\item \underline{a modification of the angular scale of the peak
    structure}: By modifying the expansion rate of the Universe at low
    redshift, the quintessence modifies the usual angular distance vs
    redshift relation (see Fig.~\ref{fig6}). This induces a global
    shift of the acoustic peak structure. Of course, if one adds a
    quintessence field while keeping fixed the Hubble parameter today,
    then this is equivalent to modifying the redshift of equivalence
    between matter and radiation, which of course has some influence
    of the acoustic peak structure (the same problem occurs for a
    cosmological constant). From Fig.~\ref{fig6}, we would have
    expected the first acoustic peak to be shifted to smaller
    multipoles since the diameter angular distance is smaller when
    $\xi > 0$. This is indeed not what is observed on
    Fig.~\ref{fign1}. This is because another, more important effect
    is that a variation of $G$ modifies the Friedmann equation at {\it
    early} times, and therefore the age of the Universe (and, hence,
    the sound horizon) is modified.  For example, when $\xi > 0$, $G$
    is larger at earlier time (see Fig.~\ref{fig0}), and the age of
    the Universe is smaller at recombination, which shifts the peak
    structure towards higher angular scales.
  
\item \underline{a boost of the Sachs-Wolfe plateau}: When the field
  starts to dominate, the density parameter of the ordinary matter
  decay, as well as the gravitational potential. This leads to the
  possibility of energy exchanges between photons and gravitational
  field, hence producing the so-called Integrated Sachs-Wolfe (ISW)
  effect which will boost the anisotropy spectrum of scales larger
  than the Hubble radius at the epoch of transition between matter and
  quintessence. The cross correlation between the Sachs-Wolfe and the
  ISW terms are difficult to compute, and, as long as the ISW term is
  not too important, one can either have a higher or lower first peak.

\end{itemize}

For any realistic model, the field is by far subdominant at
recombination. As a consequence, the acoustic peak structure of the
CMB anisotropies will not directly be affected by the field
dynamic. However, two new effects arise when one introduces a non
minimally coupled quintessence field:
\begin{itemize}

\item \underline{a modification of the amplitude of the Silk damping}:
  At small scales, viscosity and heat conduction in the photon-baryon
  fluid produce a damping of the photon
  perturbations~\cite{huwhite}. The damping scale is determined by the
  photon diffusion length at recombination, and therefore depends on
  the size of the horizon at this epoch, and hence, depends on any
  variation of the Newton constant throughout the history of the
  Universe.

\item \underline{a modification of the thickness of the last
  scattering surface}: In the same vein, the duration of recombination
  is modified by a variation of the Newton constant as the expansion
  rate is different. It is well known that CMB anisotropies are
  affected on small scales because the last scattering ``surface'' has
  a finite thickness. The net effect is to introduce an extra, roughly
  exponential, damping term, with the cutoff length being determined
  by the thickness of the last scattering
  surface~\cite{seljak94}. This thickness is of course determined by
  the duration of the recombination process. This process is
  essentially fixed in redshift units, as it is dominated by Boltzmann
  factors (and hence, by photon temperature) and not by standard
  cosmological parameters. However, when translating redshift into
  duration (or length), one has to use the Friedmann equations, which
  are affected by a variation of the Newton constant.  The relevant
  quantity to consider is the so-called visibility function $g$,
  defined as 
\begin{equation} 
  g (\eta) = \dot \tau e^{-\tau} ,
\end{equation} 
  with $\dot \tau$ being the differential opacity, and $\tau|_{\rm
  today} = 0$. In the limit of an infinitely thin last scattering
  surface, $\tau$ goes from $\infty$ to $0$ at recombination
  epoch. For standard cosmology, it drops from a large value to a much
  smaller one, and hence, the visibility function still exhibits a
  peak, but is much broader.  A few examples are given in
  Fig.~\ref{figvis} for several values of the coupling parameter
  $\xi$.  As it can be seen, the height of the visibility function
  varies with $\xi$, and therefore, its thickness also varies as the
  integral of this function is 1.
\begin{figure}[ht]
  \centerline{ \psfig{file=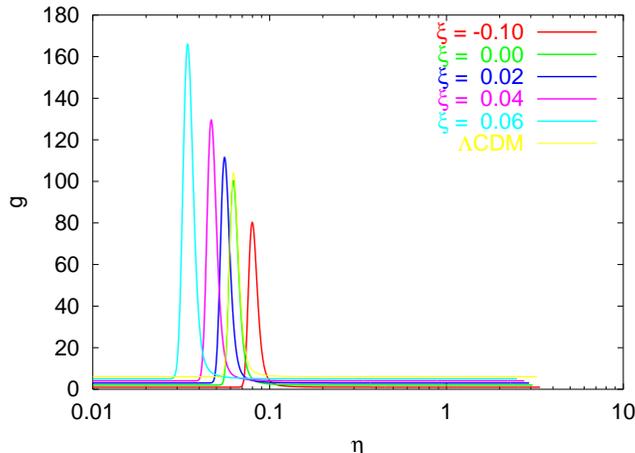,angle=270,width=3.5in}}
  \caption{Modification to the visibility function due to non minimal
  coupling between a quintessence field and curvature. Minimally
  coupled models are identical to $\Lambda$CDM models since the
  expansion history of both model is strictly identical at high
  redshift, whereas important differences can occurs when one
  introduces a non minimal coupling. For a better readability the
  various visibility functions have been slightly shifted upward.}
\label{figvis}
\end{figure}

\item \underline{a modification of the sound horizon at
recombination}: as for the photon diffusion length, the sound horizon
depends on the size of the horizon, at can significantly change if the
Newton constant is different from now at this epoch. The net effect
is to produce a shift in the Doppler peak structure.

\end{itemize}

More quantitative estimates of all these four effects are difficult to
compute analytically, since there is absolutely no reason that there
exists any simple solutions to relevant equations. However, as already
noted for the supernovae luminosity distance, the effect of the non
minimal coupling one the angular size of the acoustic peaks is small,
except for unacceptable large value of the parameters. On the
contrary, the ISW effect exhibits a much stronger dependence on $\xi$
and $B$, as shown on Fig.~\ref{fign1}. One also notices significant
modifications in the damping at high multipoles.

On Fig.~\ref{fign2}, we have plotted the corresponding matter power
spectra. Two important observable effect arise here. First, the
normalization of the spectra changes because the usual Sachs-Wolfe
formula $\delta T / T = \Phi / 3$ which relates the CMB anisotropy to
the matter power spectrum amplitudes does no longer hold because of
the ISW effect. This is particularly obvious for large, positive
values of $\xi$. Second, the maximum of the power spectrum is shifted
by the coupling. This is because this maximum give the scale that
enters into the Hubble radius at the epoch of equality between matter
and radiation. In all the cases presented here, this epoch always
corresponds to the same redshift but not to the same cosmic time
because the Friedmann equations are modified.  Therefore, the relation
$H(z)$ is different in all these models.

In practice, the power law coupling~(\ref{coupling2}) is already very
strongly constrained by the PN parameters. For all acceptable values
of the parameters, the CMB anisotropies show a negligible deviation
from the minimally coupled case (note that on Fig.~\ref{fign2}, the
plotted non zero values of $\xi$ are one order of magnitude larger
than allowed values inferred from Fig.~\ref{fig1}). On the contrary,
for the exponential coupling where the PN constraints are fulfilled by
construction, the CMB anisotropies can play a significant role in
constraining (or measuring) the coupling parameters.
\begin{figure}[ht]
  \centerline{ 
    \psfig{file=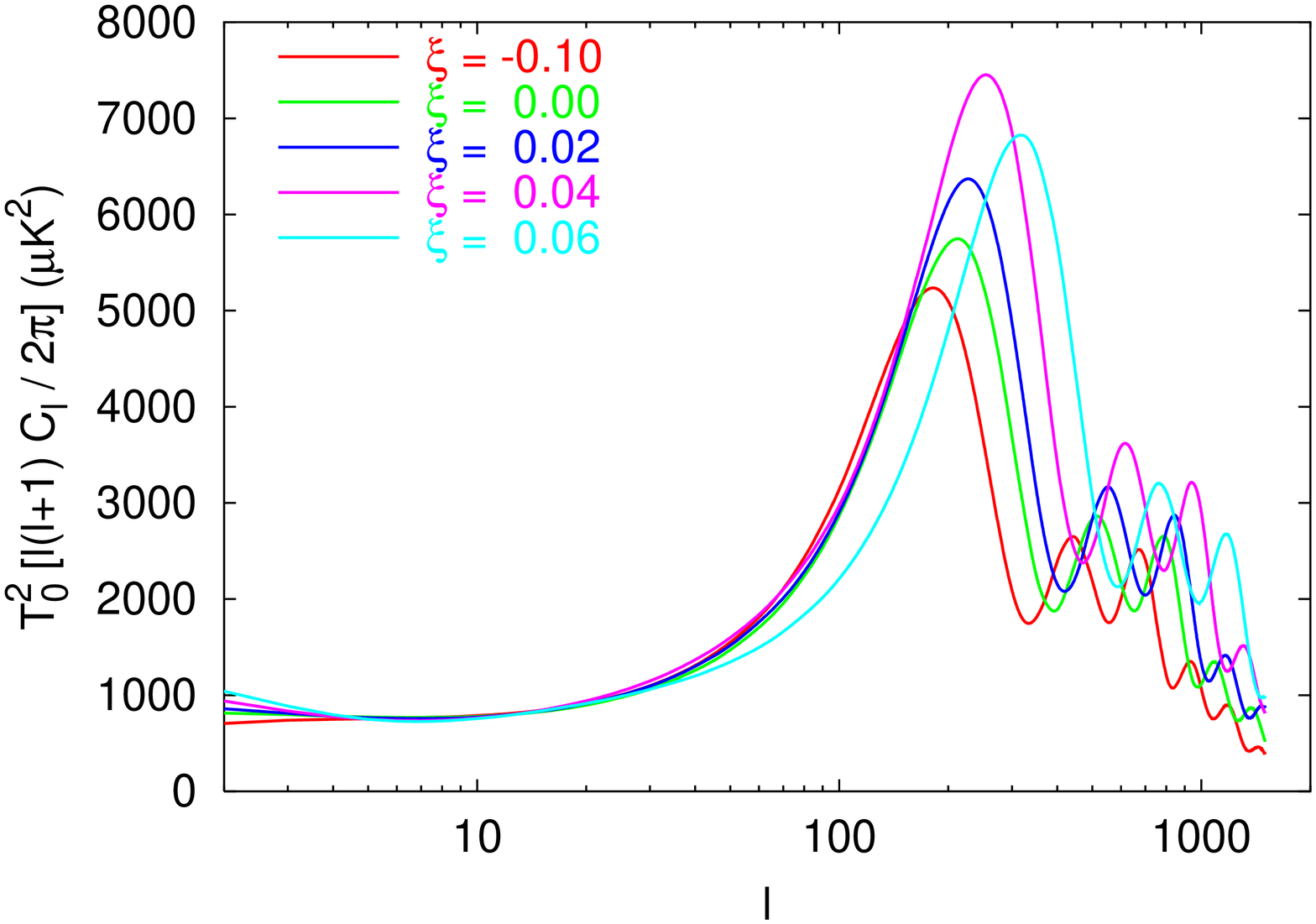,angle=270,width=3.5in}
    \psfig{file=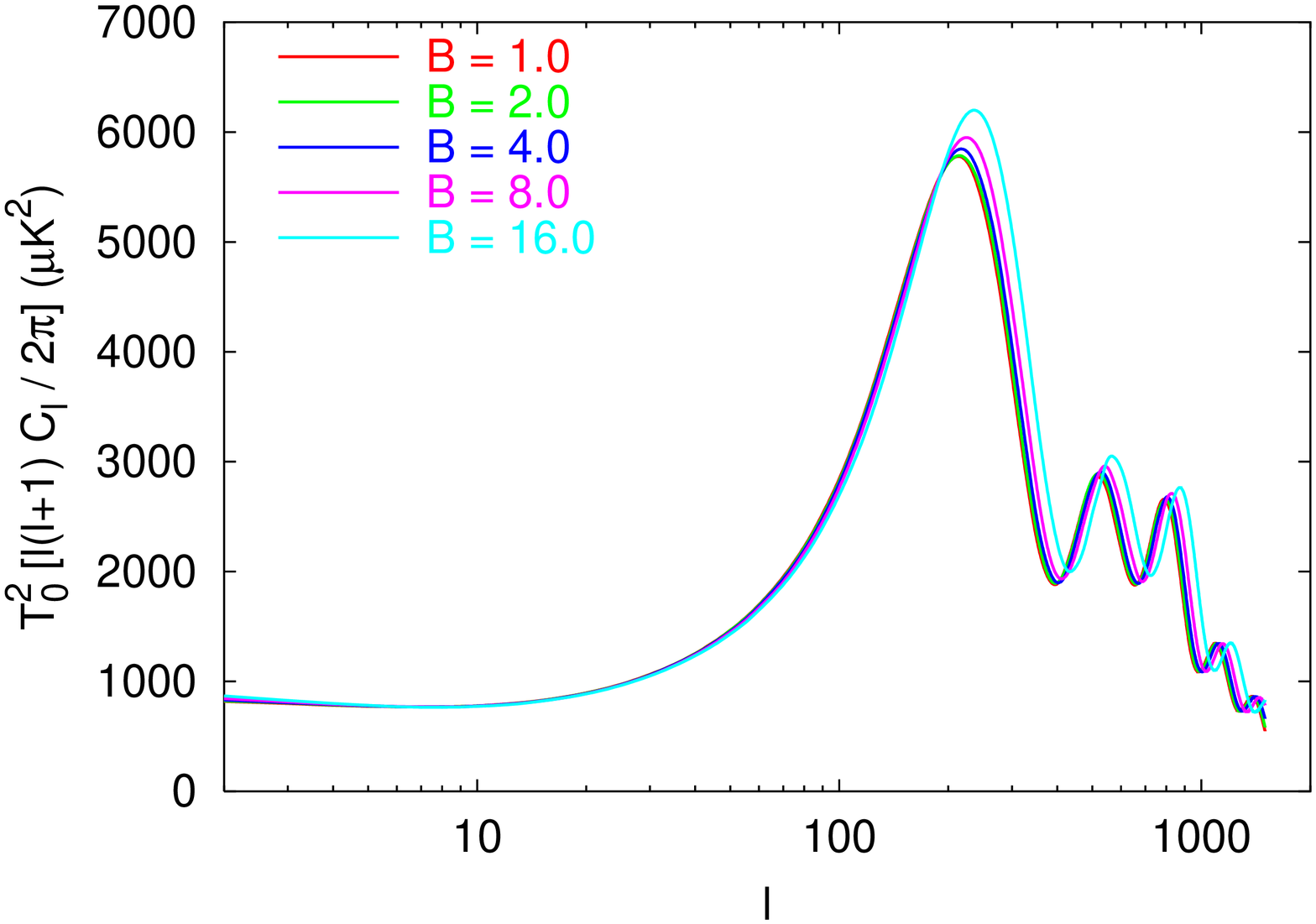,angle=270,width=3.5in}} \caption{Cosmic
    microwave background anisotropies for the SUGRA potential ($\alpha
    = 11$) with a nonminimally coupled scalar field (left) and with
    an exponential coupling (right), respectively, for different values
    of $\xi$ and $B$. We have assume a locally flat Universe. The
    strongest features are the boost of the decay of the amplitude on
    large scales due to the Integrated Sachs-Wolfe term, as well as a
    slight shift of the acoustic peaks structure to higher
    multipoles.}
\label{fign1}
\end{figure}
\begin{figure}[ht]
  \centerline{ 
    \psfig{file=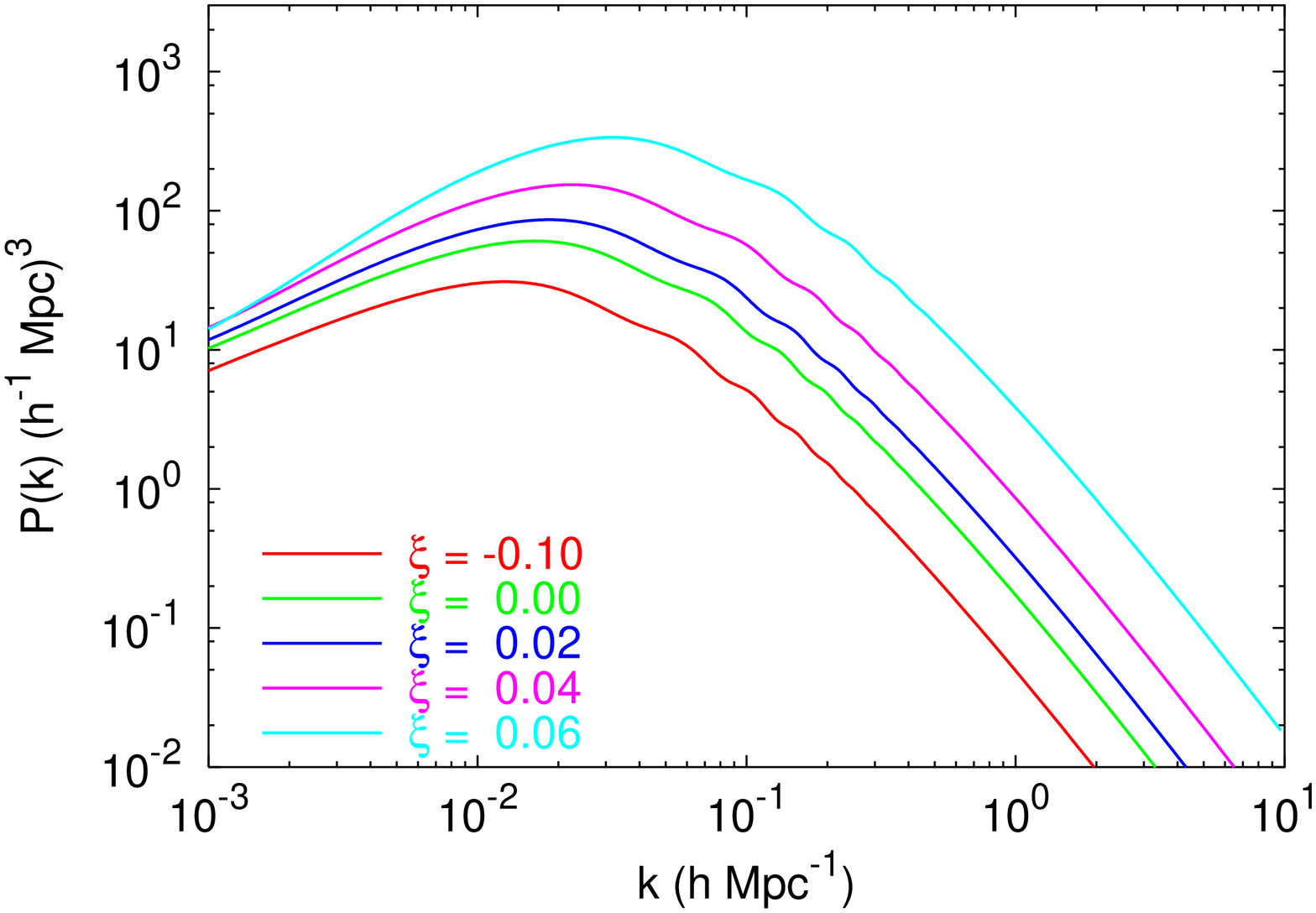,angle=270,width=3.5in}
    \psfig{file=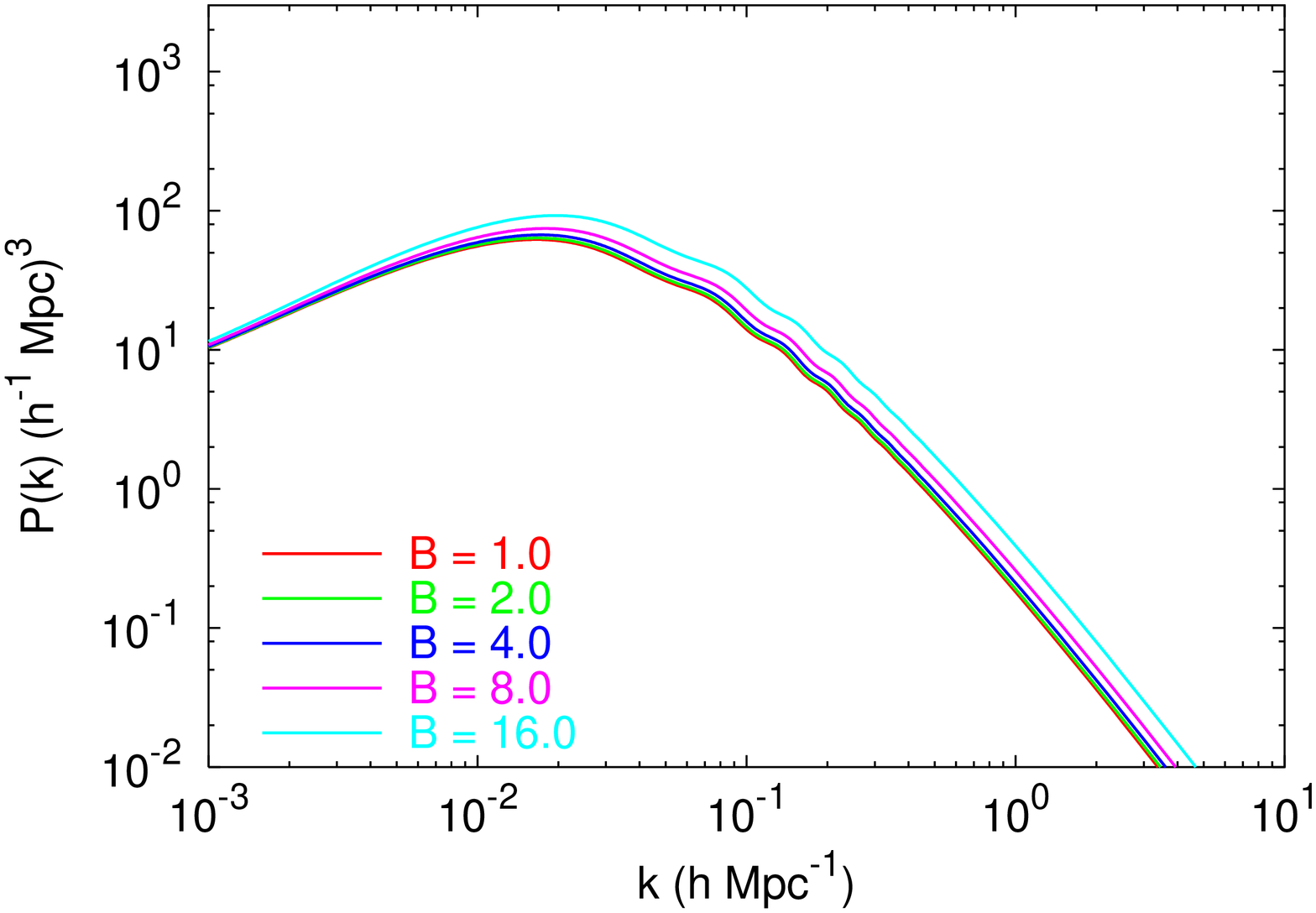,angle=270,width=3.5in}} 
    \caption{Matter power spectrum for the SUGRA potential ($\alpha =
    11$) with a nonminimally coupled scalar field (left) and with an
    exponential coupling (right), respectively, for different values of
    $\xi$ and $B$. We have assumed a locally flat Universe. The unusual
    features are the variation in the normalization of the power spectra
    once COBE-normalized, and the shift in the maximum of the spectrum
    whereas the epoch of equality is the same in all these models.}
\label{fign2}
\end{figure}

\section{Conclusion}

In this article, we have investigated the interpretation of
cosmological and astrophysical observations in the context of
scalar-tensor quintessence. In this class of models, the quintessence
field induces a time variation of the gravitational constant and the
post-Newtonian constraints and the constraints on the time variation
of the constants of nature where first imposed, as well as the
restriction arising from nucleosynthesis.

We then focused on the supernovae and cosmic microwave background
dataset. Concerning the supernovae, we extracted, using a simple toy
model, the dependence of the maximum of the light curve on the
gravitational constant. We also explained the various effects of this
coupling (mainly the modifications it induces of the Newton constant)
on CMB anisotropies and structure formation. All these features on the
CMB angular power spectrum and the matter power spectrum are
physically well understood.

All these observational constraints were applied all along the article
to two class of models: (i) a nonminimally coupled quintessence field
and (ii) and exponential coupling. We showed that the first class of
models was very constrained mainly because the deviation from the
general relativity was fixed and that the constraints were more severe
in the case of an inverse power law potential than for the SUGRA
potential. The second class seems to leave more freedom for parameters
and could possibly be constrained using the CMB.

\acknowledgements

We warmly thank Gilles Esposito-Far\`ese for numerous discussions on
scalar-tensor theories, and Robert Mochkovitch, Michel Cass\'e, Roland
Lehoucq and Christophe Balland for enlightening discussions on the
physics of supernovae. The authors acknowledge the hospitality of the
Institut d'Astrophysique de Paris where part of this work was carried
out.

\appendix

\section{Lagrangians}
\label{A}

The general action~(\ref{1}) with the ansatz~(\ref{3}) can be
decomposed as
\begin{equation}
\label{A1}
S = \int \left[  \LAGR_\GRAV
               + \LAGR_\MC
               + \LAGR_\COUPL
               + \LAGR_\MAT
         \right] \sqrt{|g|} \dd^4 x ,
\end{equation}
representing respectively the Lagrangian contribution for the gravity
(g), the minimally coupled scalar (MC) field, the coupling (co), and
the matter (mat) and given explicitly by
\begin{eqnarray}
\LAGR_\GRAV
 & = & - \frac{1}{2 \kappa} R , \\
\LAGR_\MC
 & = & \frac{1}{2} g^{\mu \nu} D_\mu \phi D_\nu \phi - V (\phi) , \\
\LAGR_\COUPL
 & = & - \xi R f (\phi) ,
\end{eqnarray}
where $D_\alpha$ is the covariant derivative of the metric
$g_{\alpha\beta}$. By varying the Lagrangian $\LAGR_\QUINT$, we obtain
the following stress-energy tensor
\begin{equation}
\label{A2}
T_{\alpha\beta}^\MC
 =   D_\alpha \phi D_\beta \phi
   - \frac{1}{2} D_\mu \phi D^\mu \phi g_{\alpha\beta}
   + V (\phi) g_{\alpha\beta} .
\end{equation}
Varying the Lagrangian $\LAGR_\COUPL$ describing the coupling, one
obtains a stress-energy tensor that can be separated into two parts
\begin{eqnarray}
\label{A3}
T_{\alpha\beta}^\KAPPA
 & = & - 2 \xi G_{\alpha\beta} f , \\
\label{A4}
T_{\alpha\beta}^\XI
 & = & - 2 \xi \left(g_{\alpha\beta} \Box - D_\alpha D_\beta \right) f ,
\end{eqnarray}
where $G_{\alpha\beta}$ is the Einstein tensor and $\Box \equiv D_\mu
D^\mu$. We have separated the coupling term into two components to
single out the part (labeled $\kappa$) which can be absorbed in a
redefinition of the Newton constant. It follows that the Einstein
equation takes the general form
\begin{equation}
\label{A5}
G_{\alpha\beta}
 = \kappa \sum_\SS{f = \MC, \XI, \KAPPA, \MAT} T^\SS{f}_{\alpha\beta} 
\end{equation}
or, equivalently, with the use of \EQN{(\ref{4})}
\begin{equation}
\label{A6}
G_{\alpha\beta}
 = \kappa_\EFF \sum_\SS{f = \MC, \XI, \MAT} T^\SS{f}_{\alpha\beta} , 
\end{equation}
with
\begin{equation}
\kappa_\EFF \equiv \frac{\kappa}{F} = \frac{\kappa}{1 + 2 \xi \kappa f}.
\end{equation}

\section{Background spacetime equations}
\label{B}

By varying the Lagrangians with respect to the scalar field, we obtain
the Klein-Gordon equation
\begin{equation}
\Box \phi + V' = - \xi R f' 
\end{equation}
or 
\begin{equation}
\ddot \phi + 2 \Hconf \dot \phi + a^2 V' = - \xi a^2 R f' ,
\end{equation}
where $R$ stands for the scalar curvature and a prime denotes a
derivative with respect to $\phi$.  Alternatively, the usual
conservation equation ($D_\mu T_\MAT^{\mu\alpha} = 0$) for the matter
has to be completed by the conservation equation of each of the
stress-energy tensors defined above
\begin{eqnarray}
\label{A7}
D_\mu T_\MC^{\mu\alpha}
 & = & - 2 \xi [\frac{R}{2}] g^{\mu \alpha} D_\mu f
       \equiv Q^\alpha_\MC , \\
\label{A8}
D_\mu T_\XI^{\mu\alpha}
 & = & - 2 \xi [- R^{\mu \alpha}] D_\mu f
       \equiv Q^\alpha_\XI , \\
\label{A9}
D_\mu T_\KAPPA^{\mu\alpha}
 & = & - 2 \xi [G^{\mu \alpha}] D_\mu f
       \equiv Q^\alpha_\KAPPA 
\end{eqnarray}
where the $Q^\nu$ can be interpreted as a force term~\cite{uzan98}.  It
is straightforward to check that, as expected, the sum of the right-hand
sides of these three equations vanishes.

When developing Eq.~(\ref{A6}) with the metric~(\ref{5}), we obtain
the Friedmann equations
\begin{eqnarray}
\label{B1}
3 (\Hconf^2 + K)
 & = & \kappa_\EFF a^2 
       \sum_\SS{f = \MAT, \MC, \XI} \rho_\SS{f} , \\
\label{B2}
2 (\Hconf^2 + K - \dot\Hconf)
 & = & - \kappa_\EFF a^2 
         \sum_\SS{f = \MAT, \MC, \XI} (P_\SS{f} + \rho_\SS{f}) ,
\end{eqnarray}
where we have introduced the quantities
\begin{eqnarray}
\label{B3}
a^2 \rho_\MC
 & \equiv & \frac{1}{2} \dot \phi^2 + a^2 V , \\
\label{B4}
a^2 P_\MC
 & \equiv & \frac{1}{2} \dot \phi^2 - a^2 V , \\
\label{B5}
a^2 \rho_\XI
 & \equiv & - 6 \xi \Hconf \dot f , \\
\label{B6}
a^2 P_\XI
 & \equiv &   2 \xi (\ddot f + \Hconf \dot f) .
\end{eqnarray}
$\rho_\MC$ and $P_\MC$ correspond to the minimally coupled part of the
scalar field energy density and pressure, $\rho_\XI$ and $P_\XI$ to
the nonminimally coupled part which can not be absorbed in
$\kappa_\EFF$. Note that these two equations are not completely
straightforward to solve since there are terms proportional to
$\Hconf$, $K$, $\dot \Hconf$ in $\rho_\XI$, $P_\XI$. For example,
defining
\begin{eqnarray}
\label{HX1}
\Hconf_\XI &\equiv & \sqrt{- K + \frac{\kappa_\EFF a^2}{3}
                       \sum_\SS{f = \MAT, \MC} \rho_\SS{f}} , \\
\epsilon_\XI &\equiv & \frac{\xi \dot f \kappa_\EFF}{\Hconf_\XI} ,
\end{eqnarray}
one gets
\begin{equation}
\Hconf = \Hconf_\XI \left[\sqrt{1 + \epsilon_\XI} - \epsilon_\XI\right] ,
\end{equation}
from which one can compute $\rho_\XI$. For $P_\XI$, one can use the
following formulas:
\begin{eqnarray}
a^2 \tilde P_\XI &\equiv & 2 \xi \left[  f'' \dot \phi^2
                                   - \Hconf f' \dot \phi 
                                   - a^2 V' f' \right] , \\
\beta &\equiv & 2 \xi^2 f' {}^2 \kappa_\EFF , \\
\label{HX2}
P_\XI & = &   \tilde P_\XI
            + \frac{\beta}{1 + 3 \beta}
              \left[  \rho_\XI - 3 \tilde P_\XI
                    + \sum_\SS{f = \MAT, \MC} (\rho_\SS{f} - 3 P_\SS{f})
              \right] .
\end{eqnarray}
We also set
\begin{eqnarray}
\label{B7}
\rho_\QUINT & \equiv & \rho_\MC + \rho_\XI , \\
P_\QUINT & \equiv & P_\MC + P_\XI ,
\end{eqnarray}
which are the contribution of the scalar field entering the right-hand
side of the Friedmann equations, and
\begin{eqnarray}
\label{B8}
a^2 \rho_\KAPPA
 & \equiv & - 6 \xi \left(\Hconf^2 + K\right) f , \\
a^2 P_\KAPPA
 & \equiv & 2 \xi \left(2 \dot \Hconf + \Hconf^2 + K \right) f ,
\end{eqnarray}
which are the quantities entering the redefinition of $\kappa$. With
these notations, the conservation equations~(\ref{A7})--(\ref{A9})
take the form
\begin{eqnarray}
\label{B9}
\dot \rho_\MC + 3 \Hconf \left(\rho_\MC + P_\MC\right)
 & = &   - \frac{6 \xi \dot f}{a^2}
         \left[\Hconf^2 + K + \dot \Hconf\right] , \\
\label{B10}
\dot \rho_\XI + 3\Hconf \left(\rho_\XI + P_\XI \right)
 & = & \frac{6 \xi \dot f}{a^2} [\dot \Hconf] , \\
\label{B11}
\dot \rho_\KAPPA + 3 \Hconf \left(\rho_\KAPPA + P_\KAPPA\right)
 & = & \frac{6 \xi \dot f}{a^2} \left[\Hconf^2 + K\right] ,
\end{eqnarray}
and it can be checked, as expected, that the sum of the right-hand
sides of \EQNS{(\ref{B9})--(\ref{B11})} vanishes.

In practice, it is not difficult to solve this set of equations [apart
from the subtleties described in \EQNS({\ref{HX1})--(\ref{HX2})}]. The
main problem arises from the fact that we know {\it a priori} neither
the values of the energy scale of the
potential~(\ref{01}),(\ref{sugra}), nor the bare Einstein constant
$\kappa$. All what we know (or impose) are the value of
$\Omega_\QUINT$ and $\kappa_\EFF$ today.  We are therefore obliged to
use standard routines~\cite{NR} to find a solution that converges
towards the desired values for $\Omega_\QUINT$ and $\kappa_\EFF$
today.  In order to do so, one is obliged to provide a reasonably good
starting point for $\kappa$ and $M$. In practice, one can consider
\begin{eqnarray}
\kappa_\START & = & 8 \pi G , \\
M_\START & = & (\rho_\CRIT M_\PL^\alpha)^\frac{1}{4 + \alpha} ,
\end{eqnarray}
which usually converges for reasonable values of the parameters. When
this starting point fails to converge, we are usually in a region of
the parameter space already ruled out by the astrophysical constraints
detailed in Sec.~\ref{IIa}.

\section{Perturbation equations}
\label{C}

We consider the general form of a perturbed cosmological spacetime
\begin{equation}\label{C1}
\dd s^2
 = a^2
   \left[  (1 + 2 A) \dd \eta^2
         + 2 (\nabla_i B + \VV{B}_i) \dd \eta \dd x^i
         - (  \gamma_{ij} + 2 C \gamma_{ij}
            + 2 \nabla_i \nabla_j E + 2 \nabla_{(i} \VV{E}_{i)}
            + 2 \TT{E}_{ij}) 
           \dd x^i \dd x^j
   \right] ,
\end{equation}
where $\gamma_{ij}$ is the metric of the homogeneous spatial sections
and $\nabla_i$ its covariant derivative. The vector
quantities $\VV{B}_i$ and $\VV{E}_i$ are divergenceless ($\nabla^i
\VV{B}_i = \nabla^i \VV{E}_i = 0$), and the tensor $\TT{E}_{ij}$ is
divergenceless and traceless.

For any fluid, the stress-energy tensor can be decomposed as
\begin{equation}
T_{\alpha \beta}
 = (P + \rho) u_\alpha u_\beta - P g_{\alpha \beta} + \Pi_{\alpha \beta} ,
\end{equation}
where $\Pi_{\alpha \beta}$ is a traceless tensor orthogonal the the
four-velocity $u^\alpha$. For a perfect fluid, we have $\Pi_{\alpha
  \beta} = 0$. We then define
\begin{eqnarray}
\label{C3b}
- a^2 (P + \rho) v_i
 & \equiv & (P + \rho) u_0 g_{ij} \delta u^j + \delta \Pi_{0i} , \\
\label{C4b}
a^2 P \pi_{ij}
 & \equiv & \delta \Pi_{ij} ,
\end{eqnarray}
where $v_i$ and $\pi_{ij}$ have to be seen as three-dimensional
quantities whose indices are raised and lowered by the metric
$\gamma_{ij}$. Using these quantities, the stress-energy tensor
perturbation takes the form
\begin{eqnarray}
\label{C2}
\delta T_{00}
 & = & \rho a^2 \left(\delta + 2 A \right) , \\
\label{C3}
\delta T_{0i}
 & = & \rho a^2 \left[  \nabla_i B + \VV{B}_i
                      - (1 + \omega) (\nabla_i v + \VV{v}_i) \right] , \\
\label{C4}
\delta T_{ij}
 & = & P a^2 \left[  h_{ij}
                   + \frac{\delta P}{P} \gamma_{ij}
                   + \Delta_{ij} \pi
                   + D_{(i} \VV{\pi}_{j)} + \TT{\pi}_{ij}
             \right] ,
\end{eqnarray}
where $h_{ij} \equiv 2 C \gamma_{ij} + 2 \nabla_i \nabla_j E + 2
\nabla_{(i} \VV{E}_{i)} + 2 \TT{E}_{ij}$, $\Delta_{ij} \equiv \nabla_i
\nabla_j - \gamma_{ij} \Delta / 3$, $\Delta \equiv \nabla_i \nabla^i$
being the Laplacian, and $\pi$, $\VV{\pi}_i$ and $\TT{\pi}_{ij}$ are
the scalar, vector and tensor components of the anisotropic stress
tensor.

We work in the gauge in which $B = E = 0$ and $\VV{E}_i = 0$ and
introduce the gauge invariant perturbation variables, labeled with a
superscript $\sharp$,
\begin{eqnarray}
\label{C5}
\delta^\sharp
 &\equiv & \frac{\delta \rho}{\rho} - \frac{\dot \rho}{\rho} (B + \dot E) , \\
\label{C6}
v^\sharp
 & \equiv & v + \dot E , \\
\label{C7}
\VV{v}^\sharp_i
 & \equiv & \VV{v}_i + \dot{\VV{E}}_i ,
\end{eqnarray}
and the four gauge invariant gravitational potentials
\begin{eqnarray}
\label{C8}
\Psi & \equiv & A + \Hconf (B + \dot E) - (\dot B + \ddot E) , \\
\Phi & \equiv & - C + \Hconf (B + \dot E) , \\
\Chi & \equiv &   A - C 
                - \frac{\partial}{\partial \eta} 
                  \left(\frac{C}{\Hconf} \right) , \\
\VV{V}_i & \equiv & \VV{B}_i +\dot{\VV{E}}_i .
\end{eqnarray}
(Note that $\Chi$ can be expressed in terms of $\Phi$ and $\Psi$ and
their time derivatives. It will, however, prove useful to work with
these three quantities.) The pressure perturbations are related to the
density contrasts by
\begin{equation}
\label{C9}
\frac{\delta P^\sharp}{P} = \frac{c_s^2}{\omega} {\delta^\sharp} + \Gamma ,
\end{equation}
where $c_s^2 = \dot P / \dot \rho$ is the (adiabatic) sound speed and
$\Gamma$ is the entropy perturbation. As we shall see later,
$\Gamma_\XI$ and $\Gamma_\MC$ do not vanish.  We also introduce the
convenient flat-slicing gauge in which $C = E = 0$ and $\VV{E}_i = 0$
(and thus where $A = \Chi$ and $B = \Phi / \Hconf$) and where the gauge
invariant density contrast and pressure perturbations, labeled with a
superscript $\flat$, are given by
\begin{eqnarray}
\label{C10}
\delta^\flat
 & \equiv &   \frac{\delta \rho}{\rho}
            - \frac{\dot \rho}{\rho} \frac{C}{\Hconf}
   =          \delta^\sharp + \frac{\dot \rho}{\rho} \frac{\Psi}{\Hconf} , \\
\frac{\delta P^\flat}{P}
 & = & \frac{\delta P^\sharp}{P} + \frac{\dot P}{P} \frac{\Psi}{\Hconf} .
\end{eqnarray}
Note that in \EQN{(\ref{C10})} we do not use the conservation equation
to express $\dot \rho$ in order for this definition to be valid also for
coupled fluids. The velocity perturbations are identical in both gauges.

The scalar field is decomposed as $\phi (\eta) + \delta \phi$ and we
introduce the two gauge invariant variables, respectively, in Newtonian
and flat-slicing gauge
\begin{eqnarray}
\label{C11}
\delta \phi^\sharp & \equiv & \delta \phi - \dot \phi (B + \dot E) , \\
\delta \phi^\flat & \equiv &   \delta \phi^\sharp
                             + \dot \phi \frac{\Phi}{\Hconf} .
\end{eqnarray}
It is indeed not the purpose of this Appendix to rederive all the
equations of perturbation; our aim is to take into account the
nonminimally coupled scalar field. For that purpose, we first compute
Klein-Gordon equation. Then, we compute the perturbed stress-energy
tensor to deduce the Einstein equations, using the standard
scalar-vector-tensor decomposition.

\subsection{Perturbed Klein-Gordon equation}

The perturbed Klein-Gordon equation reduces to
\begin{equation}
\label{pertKG}
   \ddot {\delta \phi}^\flat - \Delta \delta \phi^\flat
 + 2 \Hconf \dot {\delta \phi}^\flat + a^2 V'' \delta \phi^\flat
 - 2 \Chi \ddot \phi
 + \dot \phi \left[  \Delta \frac{\Phi}{\Hconf}
                   - \dot \Chi - 4 \Hconf \Chi \right]
 = - \xi (a^2 R f'' \delta \phi^\flat + a^2 \delta R^\flat f' ) ,
\end{equation}
where we have used the scalar curvature $R$
\begin{equation}
a^2 R = - 6 (\Hconf^2 + K + \dot \Hconf) ,
\end{equation}
and of its perturbation $\delta R^\flat$, in the flat-slicing gauge
\begin{equation}
a^2 \delta R^\flat
 =   2 \Delta (\Psi - 2 \Phi)
   + 6 (2 \Hconf^2 \Chi + 2 \dot \Hconf \Chi + \Hconf \dot \Chi) .
\end{equation}

\subsection{Perturbed fluid quantities}

After some manipulations, it appears that it is possible to arrange
the stress-energy tensor $T_{\alpha\beta}^\MC$ so that a minimally
coupled scalar field essentially behaves similar to a fluid:
\begin{eqnarray}
\rho_\MC & = & \frac{1}{2} D_\mu \phi D^\mu \phi + V , \\
P_\MC & = & \frac{1}{2} D_\mu \phi D^\mu \phi - V , \\
u_\MC^\alpha & = & \frac{D^\alpha \phi}{\sqrt{D_\mu \phi D^\mu \phi}} , \\
\Pi_\MC^{\alpha \beta} & = & 0 .
\end{eqnarray}
From these expressions, it is then possible to compute density and
pressure perturbations, etc.,
\begin{eqnarray}
\label{alain1}
a^2 \delta \rho_\MC^\flat
 & = &   \dot \phi \dot {\delta \phi}^\flat
       + a^2 V' \delta \phi^\flat + \dot \phi^2 \Chi , \\
a^2 \delta P_\MC^\flat
 & = &   \dot \phi \dot {\delta \phi}^\flat
       - a^2 V' \delta \phi^\flat + \dot \phi^2 \Chi , \\
a^2 (P_\MC + \rho_\MC) v^\flat_\MC 
 & = & - \dot \phi \delta \phi^\flat + \dot \phi^2 \frac{\Phi}{\Hconf} , \\
a^2 (P_\MC + \rho_\MC) \VV v^\flat_{\MC \; i}
 & = & \dot \phi^2 \VV{V}_i , \\
a^2 P_\MC \pi_{ij}^\MC & = & 0 .
\end{eqnarray}

Similar expressions can be found for the $\xi$ part. For the density
and the pressure, we have
\begin{eqnarray}
\rho_\XI & = & \frac{1}{4} (- 6 \xi \Box f + 3 \sqrt{Y}) , \\
P_\XI & = & \frac{1}{4} (6 \xi \Box f + \sqrt{Y}) ,
\end{eqnarray}
with
\begin{equation}
Y = \frac{1}{3} \left[  4 T^\XI_{\mu \nu} T_\XI^{\mu \nu}
                      - (g_{\mu \nu} T_\XI^{\mu \nu})^2 \right]
  = \left[\frac{2 \xi}{a^2} (\ddot f - 2 \Hconf \dot f) \right]^2 .
\end{equation}
As for the minimally coupled part, we set
\begin{equation}
u_\XI^\alpha = \frac{D^\alpha f}{\sqrt{D_\mu f D^\mu f}} ,
\end{equation}
from which we deduce an anisotropic stress $\Pi_\XI^{\alpha \beta}$: 
\begin{equation}
\Pi_\XI^{\alpha \beta}
 =   [  T_\XI^{\alpha \beta}
      - \frac{1}{4} g^{\alpha \beta} T_\XI^{\mu \nu} g_{\mu \nu}]
   - \sqrt{Y} 
     \left(u_\XI^\alpha u_\XI^\beta - \frac{1}{4} g^{\alpha \beta} \right) .
\end{equation}
One can easily check that $\Pi_\XI^{\alpha \beta} = 0$ and $\delta
\Pi_\XI^{00} = 0$. Using
Eqs.~(\ref{C3b}),(\ref{C4b}),(\ref{C3}),(\ref{C4}), we finally obtain
\begin{eqnarray}
a^2 \delta \rho_\XI^\flat
 & = & 2 \xi \left[- 3 \Hconf \dot{\delta f}^\flat
                   + 6 \Hconf \dot f X
                   + \Delta \left(  \delta f^\flat
                                 - \dot f \frac{\Phi}{\Hconf}\right)
             \right] , \\
a^2 \delta P_\XI^\flat
 & = & 2 \xi \left[  \ddot \delta f^\flat + \Hconf \dot{\delta f}^\flat
                   - \frac{2}{3} 
                     \Delta \left(  \delta f^\flat
                                  - \dot f \frac{\Phi}{\Hconf}\right)
                   - \dot f \dot \Chi - 2 \ddot f \Chi - 2 \Hconf \dot f \Chi 
             \right] , \\
a^2 (P_\XI + \rho_\XI) v^\flat_\XI 
 & = & 2 \xi \left[- \dot {\delta f}{}^\flat + \Hconf \delta f^\flat
                   + (\ddot f - 2 \Hconf \dot f) \frac{\Phi}{\Hconf}
                   +  \dot f \Chi \right] , \\
a^2 (P_\XI + \rho_\XI) \VV v^\flat_{\XI \; i}
 & = & 2 \xi (\ddot f - 2 \Hconf f) \VV{V}_i , \\
a^2 P_\XI \pi^\XI
 & = & - 2 \xi \left(\dot f \frac{\Phi}{\Hconf} - \delta f^\flat \right), \\
a^2 P_\XI \VV{\pi}_i^\XI
 & = & - 2 \xi \dot f \VV{V}_i , \\
\label{alain2}
a^2 P_\XI \TT \pi_{ij}^\XI
 & = & - 2 \xi \dot f \dot {\TT E}_{ij}.
\end{eqnarray}

Finally, the perturbation of $T_{\alpha\beta}^\KAPPA$ gives
\begin{equation}
\delta T_{\alpha\beta}^\KAPPA
 =        2 \xi \delta G_{\alpha\beta} f + 2 \xi G_{\alpha\beta} \delta f 
 \equiv   \delta T_{\alpha\beta}^{\KAPPA_\LEFT} 
        + \delta T_{\alpha\beta}^{\KAPPA_\RIGHT} .
\end{equation}
In the next section, we shall inject these results into the Einstein
equations.

\subsection{Einstein Equations}

With these quantities and the standard expressions of the perturbed
Einstein tensor, which can be found, \EG, in Ref.~\cite{pertgeom}, we
can give the perturbed Einstein equations
\begin{equation}
\delta G_{\alpha\beta}
 = \kappa \sum_\SS{f = \MAT, \KAPPA, \XI, \MC}\delta T_{\alpha\beta}^\SS{f}
\Longleftrightarrow
\delta G_{\alpha\beta}
 = \kappa_\EFF \sum_\SS{f = \MAT, \KAPPA_\RIGHT, \XI, \MC} 
               \delta T_{\alpha\beta}^\SS{f} 
\end{equation}
in the scalar-vector-tensor decomposition.

\subsubsection{Scalar modes}

After some long but straightforward manipulations, we finally obtain
\begin{eqnarray}
2 (\Delta \Phi - 3 \Hconf^2 \Chi)
 & = & 3 \Hconf^2 
       \sum _\SS{f = \MAT, \XI, \MC} 
            \Omega_\SS{f} 
            \left(  \delta^\flat_\SS{f}
                  + \frac{\delta \kappa_\EFF^\flat}{\kappa_\EFF} \right) , \\
- 2 \left[  \Hconf^2 \Chi + \left(\dot \Hconf - \Hconf^2\right) \Phi\right]
 & = & 3 \Hconf^2 \sum _\SS{f = \MAT, \XI, \MC}
                       \Omega_\SS{f} (1 + \omega_\SS{f})
                       \Hconf v^\flat_\SS{f} , \\
\Phi - \Psi
 & = & 3 \Hconf^2 \sum _\SS{f = \MAT, \XI, \MC}
                       \Omega_\SS{f} \omega_\SS{f} \pi_\SS{f} , \\
2 \left[  \Hconf^2 \Chi + 2 \dot \Hconf \Chi + \Hconf \dot \Chi
        + \frac{1}{3} \Delta (\Psi - \Phi) \right]
 & = & 3 \Hconf^2
       \sum _\SS{f = \MAT, \XI, \MC}
            \Omega_\SS{f} \omega_\SS{f} 
            \left(  \frac{\delta P^\flat_\SS{f}}{P_\SS{f}}
                  + \frac{\delta \kappa_\EFF^\flat}{\kappa_\EFF} \right) .
\end{eqnarray}
We therefore end with a linear system relating the four metric
perturbations $\Phi$, $\Psi$, $\Chi$, $\dot \Chi$ to the fluid
perturbed quantities, some of which depending on the metric
perturbations, [cf \EQNS{(\ref{alain1})--(\ref{alain2})}]. This system
can easily be solved numerically in order to extract the metric
perturbations, which in turn source the perturbed mass conservation,
and Euler and Klein-Gordon equations. Apart from the different
expressions for the metric perturbations, the other fluid conservation
equation are not modified by the presence of the quintessence
field. Hence it is quite straightforward to implement this in any
preexisting numerical code.

\subsubsection{Vector modes}

In the same way, we obtain
\begin{eqnarray}
\left[-\frac{1}{2} (\Delta + 2 K) + 2 (\Hconf^2 + K - \dot \Hconf) \right]
\VV{V}_i 
 & = & 3 \Hconf^2 \sum _\SS{f = \MAT, \XI, \MC} 
                       \Omega_\SS{f} (1 + \omega_\SS{f})
                       \Hconf v^{\SS{f} \; \flat}_i , \\
2 \Hconf^2 \VV{V}_i + \Hconf \dot{\VV{V}}_i 
 & = & 3 \Hconf^2 \sum _\SS{f = \MAT, \XI, \MC}
                       \Omega_\SS{f} \omega_\SS{f} \VV{\pi}^\SS{f}_i ,
\end{eqnarray}
which is then easily solved.

\subsubsection{Tensor modes}

The equation of evolution of the gravitational waves reads
\begin{equation}
\ddot {\TT E}_{ij} + 2 \Hconf \dot {\TT E}_{ij} + (2 K - \Delta) \TT{E}_{ij}
 = 3 \Hconf^2 \sum _\SS{f = \MAT, \XI, \MC}
                       \Omega_\SS{f} \omega_\SS{f} \TT{\pi}^\SS{f}_{ij} .
\end{equation}
The presence of the term proportional to $\dot {\TT E}_{ij}$ in
\EQN{(\ref{alain2})} indicates that the damping rate of the
gravitational waves is different. This result is identical to the one
we had obtained in an earlier work~\cite{riazuelo00}.



\begin{thebibliography}{99}

\bibitem{sndata}
  
  A.G.~Riess \ETAL, \BIBT{Observational evidence from supernovae for
  an accelerating universe and a cosmological constant} Astron.\ J.\
  {\bf 116}, 1009 (1998)\PREP{astro-ph/9806396}; S.~Perlmutter \ETAL,
  \BIBT{Discovery of a supernovae explosion at half the age of the
  universe and its cosmological implications} \nat {\bf 391}, 51
  (1998)\PREP{astro-ph/9712212}.

\bibitem{cmbdata} 
  
  P.~de Bernardis \ETAL, \BIBT{A Flat Universe from High-Resolution
  Maps of the Cosmic Microwave Background Radiation} \nat {\bf 404},
  955 (2000)\PREP{astro-ph/0004404}.

\bibitem{gldata}
  
  Y.~Mellier, Annu.\ Rev.\ Astron.\ Astrophys.\ {\bf 37}, 127 (1999).

\bibitem{weinberg}
  
  S.~Weinberg, \BIBT{The cosmological constant problem} \rmp {\bf 61},
  1 (1989).

\bibitem{binetruy00}
  
  P.~Bin\'etruy, in {\it The Early Universe}, proceedings of Les
  Houches summer school, 1999, \BIBT{Cosmological constant vs
  quintessence} [Int.\ J.\ Theor.\ Phys.\ {\bf 39} 1859
  (2000)]\PREP{hep-ph/0005037}.


\bibitem{carroll00}
  
  S.M.~Carroll, \BIBT{The cosmological constant} Living Rev.\ Relativ.\
  {\bf 4}, 1 (2001)\PREP{astro-ph/0004075}.

\bibitem{ratra}
  
  B.~Ratra and P.J.E.~Peebles, \BIBT{Cosmological consequences of a
  rolling homogeneous scalar field} \prd {\bf 37}, 3406 (1988).

\bibitem{wett}
  
  C.Wetterich, \BIBT{Cosmology and the fate of dilatation symmetry}
  Nucl.\ Phy.\ {\bf B302}, 668 (1988).

\bibitem{cds}
  
  R.R.~Caldwell, R.~Dave, and P.J.~Steinhardt, \BIBT{Cosmological
  imprint of an energy component with general equation of state} \prl
  {\bf 80}, 1582 (1998)\PREP{astro-ph/9708069}.

\bibitem{binetruy99}
  
  P.~Bin\'etruy, \BIBT{Models of dynamical supersymmetry breaking and
  quintessence} \prd {\bf 60}, 063502 (1999)\PREP{hep-th/9810553}.

\bibitem{brax99}
  
  P.~Brax and J.~Martin, \BIBT{The robustness of quintessence} \prd
  {\bf 61}, 103502 (2000)\PREP{astro-ph/9912046}.

\bibitem{zlatev}
  
  I.~Zlatev, L.~Wang, and P.J.~Steinhardt, \BIBT{Quintessence, Cosmic
  Coincidence, and the Cosmological Constant} \prl {\bf 82}, 896
  (1999)\PREP{astro-ph/9807002}.

\bibitem{riazuelo00}
  
  A.~Riazuelo and J.--P.~Uzan, \BIBT{Quintessence and Gravitational
  Waves} \prd {\bf 62}, 083506 (2000)\PREP{astro-ph/0004156}.

\bibitem{bb}

  K. Benabed and F. Bernardeau, \BIBT{Testing quintessence models with
  large-scale structure growth} \prd {\bf 64}, 083501
  (2001)\PREP{astro-ph/0104371}.


\bibitem{carollprl}
  
  S.M.~Carroll, \prl {\bf 81}, 3067 (1998).

\bibitem{brax1}
  
  P.~Brax and J.~Martin, \BIBT{Quintessence and Supergravity} \pl B
  {\bf 468}, 40 (1999)\PREP{astro-ph/9905040}.

\bibitem{gsw}
  
  M.B.~Green, J.H.~Schwarz, and E.~Witten, {\it Superstring Theory}
  (Cambridge University Press, Cambridge, England, 1987).

\bibitem{bp}
  
  N.~Banerjee and D.~Pavon, \BIBT{A quintessence scalar field in
  Brans-Dicke theory} Class.\ Quantum Grav.\ {\bf 18}, 593
  (2001)\PREP{gr-qc/0012098}.

\bibitem{uzan99}
  
  J.--P.~Uzan, \BIBT{Cosmological scaling solutions of nonminimally
  coupled scalar fields} \prd {\bf 59}, 123510
  (1999)\PREP{gr-qc/9903004}.

\bibitem{amendola99b}
  
  L.~Amendola, \BIBT{Coupled Quintessence} \prd {\bf 62}, 043511
  (2000)\PREP{astro-ph/9908023}.

\bibitem{perrotta99}
  
  F.~Perrotta, C.~Baccigalupi, and S.~Matarrese, \BIBT{Extended
  Quintessence} \prd {\bf 61}, 023507 (2000)\PREP{astro-ph/9906066}.

\bibitem{gef00b}
  
  G.~Esposito-Far\`ese and D.~Polarski, \BIBT{Scalar-tensor gravity in
  an accelerating universe} \prd {\bf 63}, 063504
  (2001)\PREP{gr-qc/0009034}.

\bibitem{will}
  
  C.M.~Will, {\it Theory and Experiments in Gravitational Physics}
  (Cambridge University Press, Cambridge, England, 1993).

\bibitem{wands94}
  
  D.~Wands, \BIBT{Extended Gravity Theories and the Einstein-Hilbert
  Action} Class.\ Quantum Grav.\ {\bf 11}, 269
  (1994)\PREP{gr-qc/9307034}.

\bibitem{damour92}
  
  T.~Damour and G.~Esposito-Far\`ese, \BIBT{Tensor multi-scalar
  theories of gravitation} Class.\ Quantum Grav.\ {\bf 9}, 2093 (1992).

\bibitem{amendola99a}
  
  L.~Amendola, \BIBT{Scaling solutions in general nonminimal coupling
  theories} \prd {\bf 60}, 043501 (1999)\PREP{astro-ph/9904120}.

\bibitem{ritis99}
  
  R.~de Ritis, A.A.~Marino, C.~Rubano, and P.~Scudellaro,
  \BIBT{Tracker fields from nonminimally coupled theory} \prd {\bf
  62}, 043506 (2000)\PREP{hep-th/9907198}.

\bibitem{bertolami99}
  
  O.~Bertolami and P.J.~Martins, \BIBT{Nonminimal coupling and
  quintessence} \prd {\bf 61}, 064007 (2000)\PREP{gr-qc/9910056}.

\bibitem{chiba99}
  
  T.~Chiba, \BIBT{Quintessence, the Gravitational Constant, and
  Gravity} \prd {\bf 60}, 083508 (1999)\PREP{gr-qc/9903094}.

\bibitem{bartolo}
  
  N.~Bartolo and M.~Pietroni, \BIBT{Scalar Tensor gravity and
  quintessence} \prd {\bf 61}, 023518 (2000)\PREP{hep-ph/9908521}.

\bibitem{DM}
  
  T.~Damour and K.~Nordtvedt, \BIBT{Tensor-scalar cosmological models
  and their relaxation toward general relativity} \prd {\bf 48}, 3436
  (1993).

\bibitem{chenk}
  
  X.~Chen and M.~Kamionkowsky, \BIBT{Cosmic Microwave Background
  Temperature and Polarization Anisotropy in Brans-Dicke Cosmology}
  \prd {\bf 60}, 104036 (1999)\PREP{astro-ph/9905368}.

\bibitem{bacci}
  
  C.~Baccigaluppi, S.~Matarrese, and F.~Perrotta, \BIBT{Tracking
  Extended Quintessence} \prd {\bf 62}, 123510
  (2000)\PREP{astro-ph/0005543}.

\bibitem{la}
  
  L.~Amendola, \BIBT{perturbations in a coupled scalar field
  cosmology} Month.\ Not.\ R.\ Astron.\ Soc.\ {\bf 312}, 521
  (2000)\PREP{astro-ph/9906073}.

\bibitem{bommamendola}
  
  L.~Amendola, \BIBT{Dark energy and the Boomerang data} \prl {\bf 86},
  196 (2001)\PREP{astro-ph/0006300}.

\bibitem{chen00}
  
  X.~Chen, R.J.~Scherrer, and G.~Steigman, \BIBT{Extended quintessence
  and the primordial helium abundance} \prd {\bf 63}, 123504
  (2001)\PREP{astro-ph0011531}.

\bibitem{brax00}
  
  P.~Brax, J.~Martin, and A.~Riazuelo, \BIBT{Exhaustive Study of
  Cosmic Microwave Background Anisotropies in Quintessential
  Scenarios} \prd {\bf 62}, 103505 (2000)\PREP{astro-ph/0005428}.

\bibitem{td}
  
 C. Will, \BIBT{The Confrontation between General Relativity and
 Experiment} Living Rev. Rel. {\bf4}, 4 (2001)\PREP{gr-qc/0103036}.


\bibitem{dickey}
  
  J.O.~Dickey \ETAL, Science {\bf 265}, 482 (1994).

\bibitem{fj}
  
  P.G.~Ferreira and M.~Joyce, \BIBT{Cosmology with a Primordial
  Scaling Field} \prd {\bf 58}, 023503 (1998)\PREP{astro-ph/9711102}.

\bibitem{pichondamour}
  
  T.~Damour and B.~Pichon, \BIBT{Big Bang nucleosynthesis and
  scalar-tensor gravity} \prd {\bf 59}, 104036
  (1999)\PREP{astro-ph/9807176}.

\bibitem{peebles93}
  
  P.J.E.~Peebles, {\it Principles of Physical Cosmology} (Princeton
  University Press, Princeton, 1993).

\bibitem{garcia99}
  
  E.~Garcia-Berro \ETAL, \BIBT{On the Evolution of Cosmological Type Ia
  Supernovae and the Gravitational Constant} \PREQ{astro-ph/9907440}.

\bibitem{arnett}
  
  D.~Arnett, {\it Supernovae and Nucleosynthesis} (Princeton
  university Press, Princeton, 1996).

\bibitem{robm}
  
  R.~Mochkovitch, in {\it Matter under Extreme Conditions}, edited by
  H.~Latal and W.~Schweiger (Springer Verlag, Berlin, 1994) p.\ 49.

\bibitem{amendola}
  
  L.~Amendola, S.~Corasaniti, and F.~Occhionero, \BIBT{Time
  variability of the gravitational constant and Type Ia supernovae}
  \PREQ{astro-ph/9907222}.

\bibitem{boisseau}
  
  B.~Boisseau, G.~Esposito-Far\`ese, D.~Polarski, and A.A.~Starobinsky,
  \prl {\bf 85}, 2236 (2000).


\bibitem{uzan98}
  
  J.--P.~Uzan, \BIBT{Dynamics of relativistic interacting gases: from
  a kinetic to a fluid description} Class.\ Quantum Grav.\ {\bf 15},
  1063 (1998)\PREP{gr-qc/9801108}.

\bibitem{bunn}
  
  E.~Bunn, in {\it The Cosmic Microwave Background}, edited by
  C.~Lineweaver \ETAL{} (Kluwer, Dordrecht, 1997) p.\ 135.

\bibitem{cmbfast}
  
  U.~Seljak and M.~Zaldarriaga, \BIBT{A line-of-sight integration
  approach to cosmic microwave background anisotropies} \apj {\bf 469},
  444 (1996)\PREP{astro-ph/9603033}.

\bibitem{huwhite}
  
  W.~Hu and M.~White, \BIBT{CMB anisotropies: total angular momentum
  method} \prd {\bf 56}, 596 (1997)\PREP{astro-ph/9702170}.

\bibitem{quintCMB}

  F.~Perrotta, C.~Baccigalupi, and S.~Matarrese, \BIBT{Extended
  quintessence} \prd {\bf 61}, 023507 (2000)\PREP{astro-ph/9906066};
  C.~Baccigalupi, S.~Matarrese and F.~Perrotta, \BIBT{Tracking
  extended quintessence} {\it ibid.} {\bf 62}, 123510
  (2000)\PREP{astro-ph/0005543}.

\bibitem{seljak94}

 U.~Seljak, \BIBT{A two fluid approximation for calculating the cosmic
 microwave background anisotropies} \apj Lett. {\bf 435}, L87
 (1994)\PREP{astro-ph/9406050}.

\bibitem{NR}
  
  W.~Press \ETAL, {\it Numerical recipes in C}, 2nd.\ ed.
  (Cambridge University Press, Cambridge, England, 1992).

\bibitem{pertgeom}
  
  H.~Kodama and M.~Sasaki, \BIBT{Cosmological perturbation theory}
  Prog.\ Theor.\ Phys.\ Suppl.\ {\bf 78}, 1 (1984); V.~Mukhanov,
  H.~Feldman and R.~Brandenberger, \BIBT{Theory of cosmological
  perturbations} Phys.\ Rep.\ {\bf 215}, 203 (1992); R.~Durrer,
  \BIBT{Gauge invariant cosmological perturbation theory: a general
  study and its application to the texture scenario of structure
  formation} Fundam.\ Cosmic\ Phys.\ {\bf 14}, 209 (1994).

\end{thebibliography}
\end{document}